\documentclass[lettersize,journal]{IEEEtran}
\usepackage{multirow}
\usepackage{graphicx}
\usepackage{makecell}
\usepackage{amsmath,amsfonts}
\usepackage{pifont}
\usepackage{algorithmic}
\usepackage{array}
\usepackage{booktabs}
\usepackage[utf8]{inputenc} 
\usepackage[colorlinks,bookmarksopen,bookmarksnumbered,citecolor=blue, linkcolor=red, urlcolor=blue]{hyperref}
\usepackage[caption=false,font=normalsize,labelfont=sf,textfont=sf]{subfig}
\usepackage{textcomp}
\usepackage{booktabs}
\usepackage{stfloats}
\usepackage{xcolor} 
\usepackage{url}
\usepackage{verbatim}
\usepackage{graphicx}
\def\BibTeX{{\rm B\kern-.05em{\sc i\kern-.025em b}\kern-.08em
    T\kern-.1667em\lower.7ex\hbox{E}\kern-.125emX}}
\usepackage{balance}
\begin{document}
\title{Point Cloud Feature Coding for Object Detection over an \\Error-Prone Cloud-Edge Collaborative System}
\author{Chongzhen Tian, Hui Yuan, \textit{Senior Member, IEEE}, Pan Zhao, Chang Sun,\\ Raouf Hamzaoui, \textit{Senior Member, IEEE}, and Sam Kwong, \textit{Fellow, IEEE}
\thanks{This work was supported in part by the National Natural Science Foundation of China under Grants 62222110, 62571303, and 62172259, the Foreign Experts Recruitment Plan of Chinese Ministry of Human Resources and Social Security under Grant H20251083, the Taishan Scholar Project of Shandong Province (tsqn202103001), the Shandong Provincial Natural Science Foundation under Grant ZR2022ZD38. (Corresponding author: Hui Yuan)}

\thanks{Chongzhen Tian, Hui Yuan, Pan Zhao and Chang Sun are with the School of Control Science and Engineering, Shandong University, Jinan 250061, China, and also with the Key Laboratory of Machine Intelligence and System Control, Ministry of Education, Jinan 250061, China (e-mail: cztian@mail.sdu.edu.cn; huiyuan@sdu.edu.cn; panz@mail.sdu.edu.cn;202220795@mail.sdu.edu.cn)

Raouf Hamzaoui is with the School of Engineering, Infrastructure, and Sustainability, De Montfort University, LE1 9BH Leicester, U.K. (e-mail: rhamzaoui@dmu.ac.uk).

Sam Kwong is with the School of Data Science, Lingnan University, Hong Kong (e-mail: samkwong@ln.edu.hk).
}
}

\markboth{Journal of \LaTeX\ Class Files,~Vol.~18, No.~9, September~2020}%
{How to Use the IEEEtran \LaTeX \ Templates}

\maketitle

\begin{abstract}
Cloud-edge collaboration enhances machine perception by combining the strengths of edge and cloud computing. Edge devices capture raw data (e.g., 3D point clouds) and extract salient features, which are sent to the cloud for deeper analysis and data fusion. However, efficiently and reliably transmitting features between cloud and edge devices remains a challenging problem. We focus on point cloud-based object detection and propose a task-driven point cloud compression and reliable transmission framework based on source and channel coding. To meet the low-latency and low-power requirements of edge devices, we design a lightweight yet effective feature compaction module that compresses the deepest feature among multi-scale representations by removing task-irrelevant regions and applying channel-wise dimensionality reduction to task-relevant areas. Then, a signal-to-noise ratio (SNR)-adaptive channel encoder dynamically encodes the attribute information of the compacted features, while a Low-Density Parity-Check (LDPC) encoder ensures reliable transmission of geometric information. At the cloud side, an SNR-adaptive channel decoder guides the decoding of attribute information, and the LDPC decoder corrects geometry errors. Finally, a feature decompaction module restores the channel-wise dimensionality, and a diffusion-based feature upsampling module reconstructs shallow-layer features, enabling multi-scale feature reconstruction. On the KITTI dataset, our method achieved a 172-fold reduction in feature size with 3D average precision scores of 93.17\%, 86.96\%, and 77.25\% for easy, moderate, and hard objects, respectively, over a 0 dB SNR wireless channel. Our source code will be released on GitHub at: {\color{pink} \url{https://github.com/yuanhui0325/T-PCFC}}.
\end{abstract}


\section{Introduction}
\IEEEPARstart{I}{n} recent years, with improvements in hardware computing capabilities and algorithmic perception accuracy, machine perception systems have gradually begun replacing human vision systems in various fields, including intelligent surveillance, deep-sea exploration, and autonomous driving \cite{ref1}. Among these applications, object detection has a central task, enabling machines to autonomously identify, localize, and interpret critical objects\cite{ref2}. To achieve robust performance in challenging scenarios involving complex lighting, textureless surfaces, and large depth variations, 3D point cloud-based perception algorithms are increasingly adopted over traditional 2D image-based methods \cite{ref3}. However, despite these advantages, 3D point clouds involve significantly larger data volumes, posing critical challenges for storage and transmission \cite{ref4}.

Moreover, cloud–edge collaboration has emerged as an effective framework for enhancing machine perception systems and enabling more efficient allocation of computational resources \cite{ref5}. In this framework, edge devices capture raw data and apply preliminary processing to extract task-relevant information. The resulting features are then transmitted to cloud servers for in-depth analysis to generate final perception results. However, wireless links between cloud and edge nodes are highly susceptible to channel interference \cite{ref6}, which can severely compromise data transmission reliability and consequently degrade the performance of detection algorithms. In addition, task-relevant feature point clouds typically consist of multi-scale representations \cite{ref7}, further increasing the demands on data transmission and storage. Therefore, developing efficient and reliable feature coding and transmission schemes is essential to ensure the robust operation of edge-cloud collaborative machine vision systems. 

Typically, source coding reduces data redundancy to improve communication efficiency, whereas channel coding enhances data robustness against transmission errors. For source coding, the Moving Picture Experts Group (MPEG) has introduced two point cloud compression methods: Geometry-based Point Cloud Compression (G-PCC) \cite{ref8} and Video-based Point Cloud Compression (V-PCC) \cite{ref9}. In addition to these traditional approaches, deep learning-based compression methods have been proposed to learn compact point cloud representations through data-driven optimization, achieving higher coding efficiency \cite{ref10} \cite{ref11}. A point cloud is a collection of spatial points, each associated with positional coordinates and attribute information. The attributes are typically simple measurements such as reflectance or intensity obtained from sensors. After convolutional or other feature-extraction processing, the attribute dimensions of each point expand to encode higher-level semantic or geometric information. Despite this increase in attribute dimensionality, the representation format remains the same: sparse tensors composed of coordinates and attributes. To differentiate between these two forms, we call the sensor-derived data as raw point clouds and the processed representations as feature point clouds. However, compression methods developed for raw point clouds are not well suited for feature point clouds because of their distinct data distributions \cite{ref12}. In our previous work \cite{ref5}, we proposed a transformer-based feature point cloud compression method. However, self-attention calculations on point clouds require significant computational resources. As a result, the method is not suitable for edge devices with limited capacity.

To enhance the reliability of communication systems, many channel coding methods have been proposed. Early approaches typically optimized source and channel coding independently, which often failed to achieve optimal system performance \cite{ref13}\cite{ref14}. In contrast, Joint Source–Channel Coding (JSCC) offers both theoretical and practical advantages by jointly optimizing compression and error protection \cite{ref_add1}. Modern JSCC methods, particularly those based on deep learning, have shown superior performance compared to traditional separate coding and classical JSCC schemes \cite{ref6, ref15}. While such methods are generally adequate for raw point clouds intended for human visual systems, where small reconstruction errors have little impact on subjective perception, they are insufficient for feature point clouds used in machine perception. In feature point clouds, even small errors in geometric coordinates can be significantly amplified when mapped back to the original scale, resulting in substantial degradation of detection accuracy. This observation highlights the need for an unequal protection strategy that accounts for the different error characteristics of geometric and attribute information in feature point clouds.

To address the aforementioned challenges, we propose an object detection-driven point cloud compression and reliable transmission framework. To the best of our knowledge, this is the first work that simultaneously considers both source coding and channel coding requirements in point cloud-based machine perception scenarios. Experimental results show that the proposed source coding scheme achieves effective feature compression while maintaining or improving detection accuracy. When combined with the channel coding module, the end-to-end JSCC system enables reliable transmission over noisy wireless links and outperforms direct feature transmission. The main contributions of this paper are summarized as follows.

\textbf{1) Efficient Feature Point Cloud Coding:} We propose a U-Net-based geometry compaction block, guided by a novel point-wise loss function, to effectively separate task-relevant from task-irrelevant regions. In parallel, we introduce a channel attention and sparse convolution-based attribute compaction block to remove redundant information and extract intrinsic feature representations. Together, these blocks not only enable efficient feature compression but also enhance feature quality, leading to improved task performance compared with the original detection model.

\textbf{2) Reliable Transmission with Unequal Protection:} We develop an unequal protection strategy to address the differing noise sensitivities of geometric and attribute information. For geometric information, we use an LDPC code to ensure high-fidelity transmission. For attribute information, we propose a deep learning-based JSCC method that adaptively compresses and recovers the attributes of the feature point cloud under varying wireless conditions, which improves transmission robustness. Compared with directly transmitting compressed features, the proposed approach significantly enhances feature recovery quality, especially in low-SNR scenarios.

\textbf{3) Precise Feature Reconstruction:} We introduce a diffusion model-based feature point cloud upsampling module, supported by a prompt generation block, to reconstruct multi-scale feature point clouds at the cloud device. By generating high-resolution features from compact representations, this approach eliminates the need to transmit full-scale data, reducing bandwidth consumption while preserving task accuracy and enabling efficient multi-scale reconstruction.

The remainder of this paper is organized as follows. Section II reviews related works. Section III describes the proposed method. Section IV presents the experimental results. Finally, Section V presents the conclusions and suggests future work. 

\section{Related Works}
\subsection{Point Cloud-based Object Detection}
Based on architectural design, mainstream object detection methods can be classified into two categories: one-stage and two-stage approaches \cite{ref16}. One-stage methods directly predict object categories and bounding boxes without generating explicit region proposals, generally offering higher efficiency. For instance, Zhou {\textit{et al}}. \cite{ref17} introduced the first end-to-end deep learning framework for 3D object detection. Yan \textit{et al}. \cite{ref18} further improved efficiency by applying sparse 3D convolutional networks to voxelized point clouds. Lang \textit{et al}. \cite{ref19} proposed a pillar-based representation that converts points into vertical pillars processed with 2D convolutions, enabling real-time detection. He \textit{et al}. \cite{ref20} introduced a point-wise supervision mechanism for fine-grained geometric modeling through per-point feature optimization and structure-aware loss functions. 

Two-stage methods use a cascaded architecture that decomposes object detection into region proposal generation and object localization, typically achieving higher accuracy. Point-RCNN \cite{ref21} directly processes raw point clouds to deliver high localization accuracy, and PV-RCNN \cite{ref22} advances this approach with a hybrid voxel-point architecture for richer feature learning. Voxel R-CNN \cite{ref23} simplifies the approach by generating high-quality proposals from 3D sparse convolutional features and refining them through voxel region of interest (RoI) pooling, showing that voxel-based operations can match point-based accuracy while being more computationally efficient. More recently, Wu \textit{et al}. \cite{ref24} proposed a multimodal method, VirConv-L, which combines virtual point clouds derived from images with real point clouds to enhance scene representation and perception performance.

In recent years, transformer-based and anchor-free architectures have become prominent directions in object detection research. 3DETR \cite{ref25} uses self-attention to capture long-range dependencies among all points, overcoming the local receptive field limitations of traditional convolutions. CenterNet3D \cite{ref26}, an anchor-free method, eliminates the need for manually designed anchors but faces challenges such as poor detection of small objects, sensitivity to point cloud sparsity, and limited adaptability to rotated objects. 

In summary, most 3D object detection methods rely on voxelized features, which provide structured representations of sparse point clouds and enable efficient convolutional processing. Our method is designed to exploit such features for compression, making it broadly compatible with existing voxel-based detectors and easily extensible to other voxel-oriented approaches. In our experiments, we use VirConv-L \cite{ref24} as the baseline. This lightweight method combines the advantages of two-stage detection while supporting multimodal inputs from both images and point clouds, making it well suited for real-world applications.
\vspace{-0.3cm} 
\subsection{Point Cloud Compression}
Early point cloud compression methods were primarily designed for human vision systems, aiming to optimize the trade-off between bitrate and reconstruction quality. For example, G-PCC \cite{ref8} compresses point cloud geometry using octree \cite{ref27}, predictive tree \cite{ref28}, or trisoup \cite{ref29}, while attribute compression is achieved via region-adaptive hierarchical transform \cite{ref30}, predictive transform, or lifting transform \cite{ref31}. V-PCC \cite{ref9} first projects 3D point clouds into 2D video frames and then compresses them using video coding techniques \cite{ref32}\cite{ref33}. Beyond these traditional methods, many deep learning-based methods have been proposed to achieve better compression performance. For instance, Sun \textit{et al}. \cite{ref11} developed a context-enhanced model for geometry compression. Mao \textit{et al}. \cite{ref4} proposed a frequency-based sampling network with a multi-layer progressive encoder-decoder to compress attributes. Wang \textit{et al}. introduced a versatile multiscale conditional coding framework for both geometry \cite{ref34} and attribute \cite{ref35} compression. 

However, directly applying human-centric compression methods to machine vision tasks presents several limitations. First, perceptual metrics commonly used in human-focused compression (e.g., PSNR, SSIM) often correlate weakly with machine vision performance metrics. Second, human-centric algorithms typically follow a compress-then-analyze (CTA) \cite{ref36} pipeline, which requires reconstructing the full point cloud before downstream task analysis. In contrast, the analyze-then-compress (ATC) \cite{ref36} approach, CTA generally achieves lower compression efficiency and higher computational cost.

Research on machine perception-oriented point cloud compression is relatively rare. Liu \textit{et al}. \cite{ref37} proposed a dual-branch compression framework for both human and machine perception, where the machine branch uses a point selection module to generate sparse point clouds and specialized classifiers or detectors to preserve task accuracy. Ma \textit{et al}. \cite{ref38} developed a human-machine balanced geometry compression framework that integrates semantic mining to satisfy human and machine perception requirements. Ulhaq \textit{et al}. \cite{ref39} introduced a scalable compression framework in which the base bitstream supports machine tasks while the full bitstream serves human viewing. Liang \textit{et al}. \cite{ref40} proposed RoI-guided geometry compression with strong supervision to ensure object-region quality. Gao \textit{et al}. \cite{ref41} used bird’s eye view (BEV) representations combined with feature reconstruction, bitrate, and task-related losses to enhance task performance, although BEV projection leads to height information loss. In our previous work \cite{ref5}, we proposed a 3D feature compression method for cloud-edge multimodal object detection. However, the self-attention computation on the edge side required substantial processing resources and did not consider the impact of poor channel quality on downstream task accuracy.

In summary, existing machine perception-oriented compression methods have the following limitations: (a) most require reconstructing the full raw point cloud before analysis, which consumes more computational resources than directly compressing task-relevant features; and (b) they do not consider transmission-induced distortions, which significantly reduces reliability in practical deployments.
\vspace{-0.3cm} 
\subsection{Joint Source Channel Coding}
Source coding is primarily designed to improve communication efficiency, while channel coding focuses on ensuring transmission reliability. Conventional approaches usually treat these two components independently, which can lead to suboptimal overall performance \cite{ref42}. To address this limitation, recent studies have explored deep learning-based JSCC (DeepJSCC) \cite{ref43}. For example, Yang \textit{et al}. \cite{ref44} proposed a policy network that adapts to image content and channel conditions to optimize the tradeoff between rate and quality. Xu \textit{et al}. \cite{ref45} introduced a channel-wise soft attention module for SNR-adaptive feature scaling. Wang \textit{et al}. \cite{ref46} developed an end-to-end semantic transmission framework for videos, combining nonlinear transformations with DeepJSCC.

Compared with 2D images and videos, point clouds are inherently irregular, unordered, and sparse, which makes them fundamentally different in terms of representation and error propagation \cite{ref47}. Therefore, directly extending image- or video-oriented JSCC schemes to point clouds is insufficient. In the context of point cloud transmission, Zhang \textit{et al}. \cite{ref48} designed a progressive resampling framework with DeepJSCC for adaptive bandwidth allocation, ensuring high-quality geometry reconstruction. Liu \textit{et al}. \cite{ref49} proposed a rate allocation strategy to filter semantically irrelevant information and demonstrated that DeepJSCC can effectively mitigate channel noise. However, existing point cloud DeepJSCC methods mainly focus on the compression and reconstruction of raw geometry while overlooking attributes, which are essential for perception tasks. Moreover, most of these methods are designed for human-centric perceptual metrics, limiting their applicability to machine vision scenarios. 

Therefore, we propose a task-driven feature point cloud transmission method with unequal protection. We apply LDPC channel coding to geometry information, which is highly sensitive to distortion and requires reliable transmission. For attribute information, we use a deep learning-based approach to ensure transmission efficiency. This design combines the robustness of traditional coding with the adaptability of learning-based methods, achieving a better balance between efficiency and reliability in feature point cloud transmission.

\section{Proposed Method}
We propose a feature point cloud coding method for object detection. We first describe the pipeline of the underlying object detection network. We then provide an overview of the proposed method, followed by a detailed discussion of the architectures of the proposed modules. Finally, we introduce the loss functions used.
\vspace{-0.3cm} 
\subsection{Pipeline of the Used Object Detection Network}
\begin{figure*}
    \centering
    \setlength{\abovecaptionskip}{-0.1cm}
    \includegraphics[width=1\linewidth]{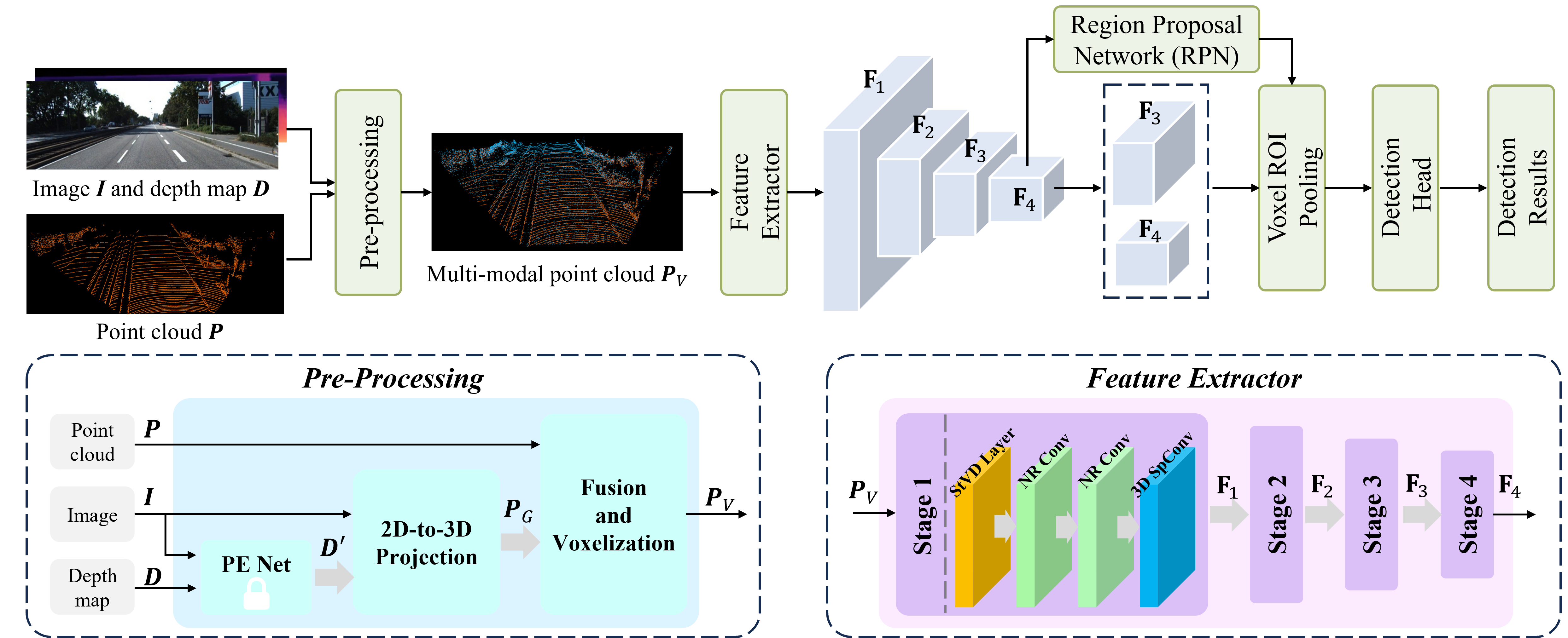}
    \caption{Pipeline of the used object detection network. StVD denotes Stochastic Voxel Discard, NR Conv denotes Noise-Resistant Submanifold Convolution and 3D SpConv denotes 3D Sparse Convolution.}
    \label{Fig.1}
\end{figure*}

To integrate visual and spatial information for object detection, the proposed system fuses data from various sensing modalities. As shown in Fig.\ref{Fig.1}, image $\textbf{\textit{I}} \in \mathbb{R}^{n_1 \times 3}$ (each row containing the $R, G, B$ values of a pixel), a depth map $\textit{\textbf{D}} \in \mathbb{R}^{n_1 \times 1}$, and a point cloud $\textit{\textbf{P}} \in \mathbb{R}^{n_2 \times 4}$ (each point represented by 3D coordinates $X, Y, Z$ and reflectance $r$) are jointly processed to generate a multi-modal point cloud. Here, $n_1$ and $n_2$ denote the numbers of pixels and points, respectively. The image $\textit{\textbf{I}}$ and depth map $\textit{\textbf{D}}$ are first processed by the pre-trained PE-Net \cite{ref50} to obtain a refined depth map $\textit{\textbf{D}}' \in \mathbb{R}^{n_1 \times 1}$. Next, $\textit{\textbf{D}}'$ and $\textit{\textbf{I}}$ are passed through a 2D-to-3D projection block to generate a virtual point cloud $\textit{\textbf{P}}_G \in \mathbb{R}^{n_1 \times 6}$, where each point is represented by its 3D coordinates and corresponding color values. The virtual point cloud $\textit{\textbf{P}}_G$ is then combined with the original point cloud $\textit{\textbf{P}}$ to form a multi-modal point cloud $\textit{\textbf{P}}_M \in \mathbb{R}^{n \times 8}$, in which each point is represented as $(X, Y, Z, R, G, B, r, f).$
For points originating from $\textbf{\textit{P}}_G$, the flag $f$ is set to $1$ and the reflectance $r$ is set to $0$. For points from $\textit{\textbf{P}}$, the color components $R, G, B$ are set to $0$ and $f$ is set to $2$. Subsequently, $\textit{\textbf{P}}_M$ is voxelized to obtain the voxelized point cloud $\textit{\textbf{P}}_V \in \mathbb{R}^{n_3 \times 11}$, where $n_3$ denotes the number of non-empty voxels. The first three columns represent the spatial coordinates of the occupied voxels, while the remaining columns store the averaged per-voxel attribute values corresponding to the eight features $(X, Y, Z, R, G, B, r, f)$ of the points contained within each voxel.

The voxelized point cloud $\textbf{P}_V$ is then fed into a feature extractor to obtain multi-scale feature tensors, denoted as $\textbf{F}_1 \in \mathbb{R}^{n \times 19}, \quad 
\textbf{F}_2 \in \mathbb{R}^{n \times 35}, \quad 
\textbf{F}_3 \in \mathbb{R}^{n \times 67}, \quad 
\textbf{F}_4 \in \mathbb{R}^{n \times 67}.$
In each feature tensor, the first three columns correspond to voxel coordinates, while the remaining columns represent the extracted feature attributes. The feature extractor consists of four stages, each comprising a \textit{stochastic voxel discard} layer, two \textit{noise-resistant submanifold convolution} layers, and a \textit{3D sparse convolution} layer. $\textbf{F}_4$ is fed into a region proposal network (RPN) to generate a set of 3D Regions of Interest (3D RoIs). Then, $\textbf{F}_3$, $\textbf{F}_4$, and the 3D RoIs are passed through a Voxel RoI pooling module, which aggregates the sparse voxel features from $\textbf{F}_3$ and $\textbf{F}_4$ within each 3D RoI to generate fixed-size representations. 

Finally, the aggregated features are fed into the detection head, which consists of shared fully connected (FC) layers followed by task-specific branches for classification and regression. The classification branch outputs category probabilities for each RoI, while the regression branch predicts center offsets, object dimensions, and yaw angles. More details about this detector can be found in \cite{ref24}.
\vspace{-0.3cm} 
\subsection{Overview of the Proposed Network}

\begin{figure*}
    \centering
    \setlength{\abovecaptionskip}{-0.1cm}
    \includegraphics[width=1\linewidth]{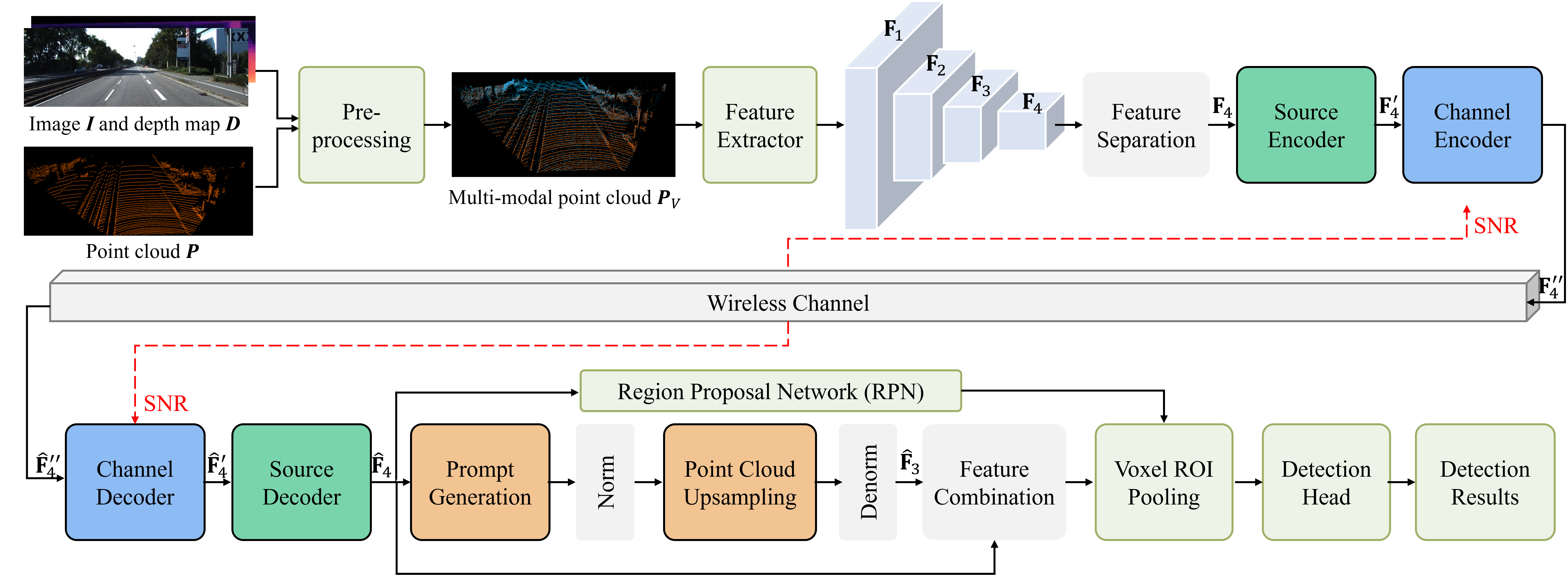}
    \caption{Overview of the proposed network.}
    \label{Fig.2}
\end{figure*}

As shown in Fig. \ref{Fig.2}, the proposed method splits the object detection network after the feature extractor and inserts a \textit{source encoder}, \textit{channel encoder}, \textit{channel decoder}, and \textit{source decoder} for efficient and reliable transmission of $\textbf{F}_4$. Considering that the remaining detection network requires multi-scale features for voxel RoI pooling, we introduce a diffusion model-based feature point cloud upsampling module to generate the high-scale feature $\hat{\textbf{F}}_3$. The details of these modules are presented in the following sections.
\vspace{-0.3cm} 
\subsection{Source Encoder}

\begin{figure}
    \centering
    \setlength{\abovecaptionskip}{-0.1cm}
    \includegraphics[width=0.9\linewidth]{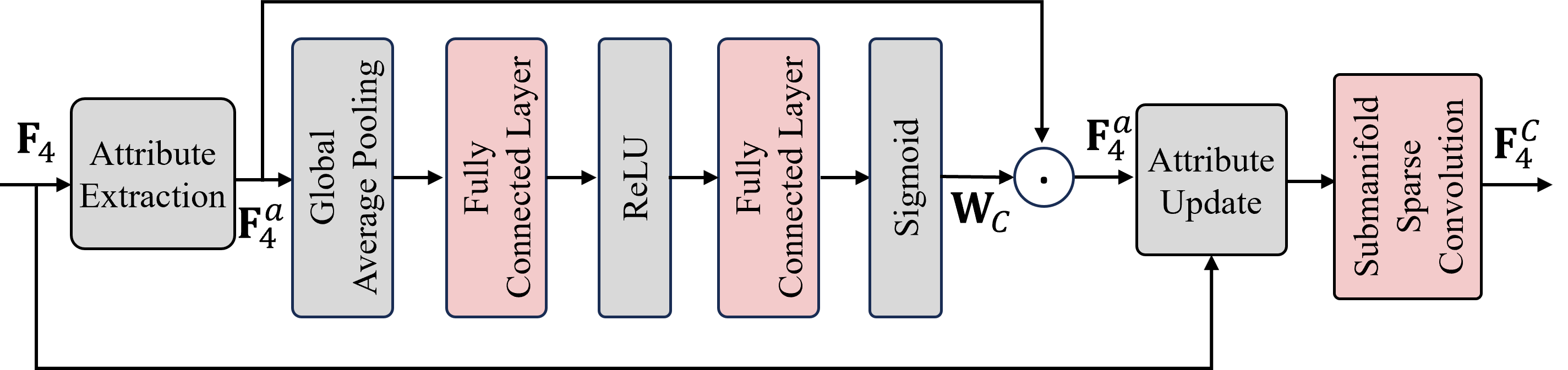}
    \caption{Channel compaction block.}
    \label{Fig.3}
\end{figure}

\textbf{Channel Compaction:} Fig.\ref{Fig.3} illustrates the channel compaction block. First, we extract the attribute information of $\textbf{F}_4$ and obtain $\textbf{F}_4^a \in \mathbb{R}^{n \times 64}$. Then, a squeeze-and-excitation (SE) attention mechanism is applied to $\textbf{F}_4^a$ to enhance feature discrimination. Specifically, we use global average pooling (GAP) on $\textbf{F}_4^a$ to obtain channel-wise descriptors. After that, two FC layers followed by ReLU (introducing nonlinearity) and sigmoid (normalizing the weights to the range $[0,1]$) activation functions, denoted as $\mathcal{FC}_R$ and $\mathcal{FC}_S$, are used to generate channel weights. The above operations can be formulated as
\begin{equation}
\textbf{W}_C = \mathcal{FC}_S \big( \mathcal{FC}_R ( \mathcal{GAP}(\textbf{F}_4^a) ) \big),
\end{equation}
where $\textbf{W}_C \in \mathbb{R}^{1 \times 64}$ denotes the generated weight vector, and $\mathcal{GAP}$ denotes the global average pooling operation. The output channel dimensions of $\mathcal{FC}_R$ and $\mathcal{FC}_S$ are set to $8$ and $64$, respectively. Then, $\textbf{W}_C$ is multiplied with $\textbf{F}_4^a$ to obtain the enhanced features. Finally, the enhanced feature is fed into a submanifold sparse convolution layer ($\mathcal{C}_S$) to achieve channel reduction. This process can be expressed as
\begin{equation}
\textbf{F}_4^C = \mathcal{C}_S \big( \textbf{W}_C \cdot \textbf{F}_4^a \big),
\end{equation}
where $\textbf{F}_4^C  \in \mathbb{R}^{n \times 9}$ denotes the channel-compressed representation of $\textbf{F}_4$. The kernel size and output channel of $\mathcal{C}_S$ are set to $1 \times 1 \times 1$ and $6$, respectively. 

\begin{figure}
    \centering
    \setlength{\abovecaptionskip}{-0.1cm}
    \includegraphics[width=1\linewidth]{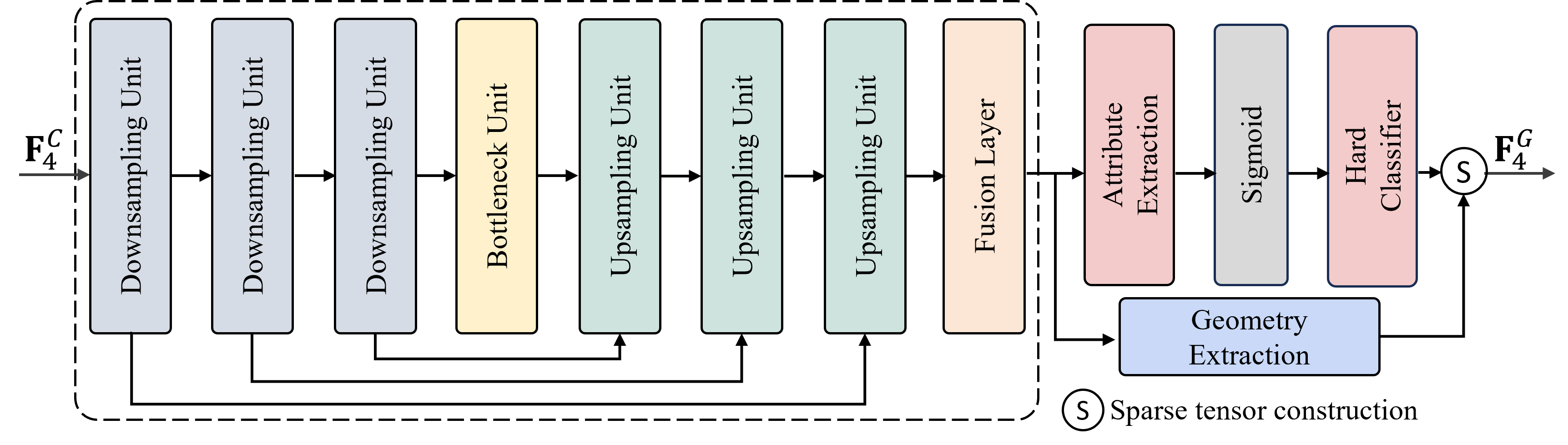}
    \caption{\centering Spatial compaction block.}
    \label{Fig.4}
\end{figure}

\textbf{Spatial Compaction:} In 3D object detection, object-relevant regions typically occupy only a small portion of the scene, while the majority corresponds to background or object-irrelevant areas. Distinguishing these regions at the edge would allow the removal of irrelevant data and enable efficient feature transmission. However, since our framework uses a feature compression pipeline that encodes the entire scene into a compact global representation, only a holistic feature map is available at the edge, without explicit separation between object-relevant and object-irrelevant regions. To address this problem, we introduce a U-Net-based spatial compaction block guided by a voxel-wise mask loss. Although this block predicts a binary mask for each voxel, the computational overhead remains low because the spatial size of $\textbf{F}_4^C$ is only 1/64 that of the original point cloud. Moreover, compared with the raw input, feature representations contain richer semantic cues, enabling accurate binary classification with only a few additional layers. An illustration of the spatial compaction block is provided in Fig. \ref{Fig.4}, and the details of the proposed loss function and block architecture are presented in the following sections.

\begin{figure}
    \centering
    \setlength{\abovecaptionskip}{-0.1cm}
    \includegraphics[width=1\linewidth]{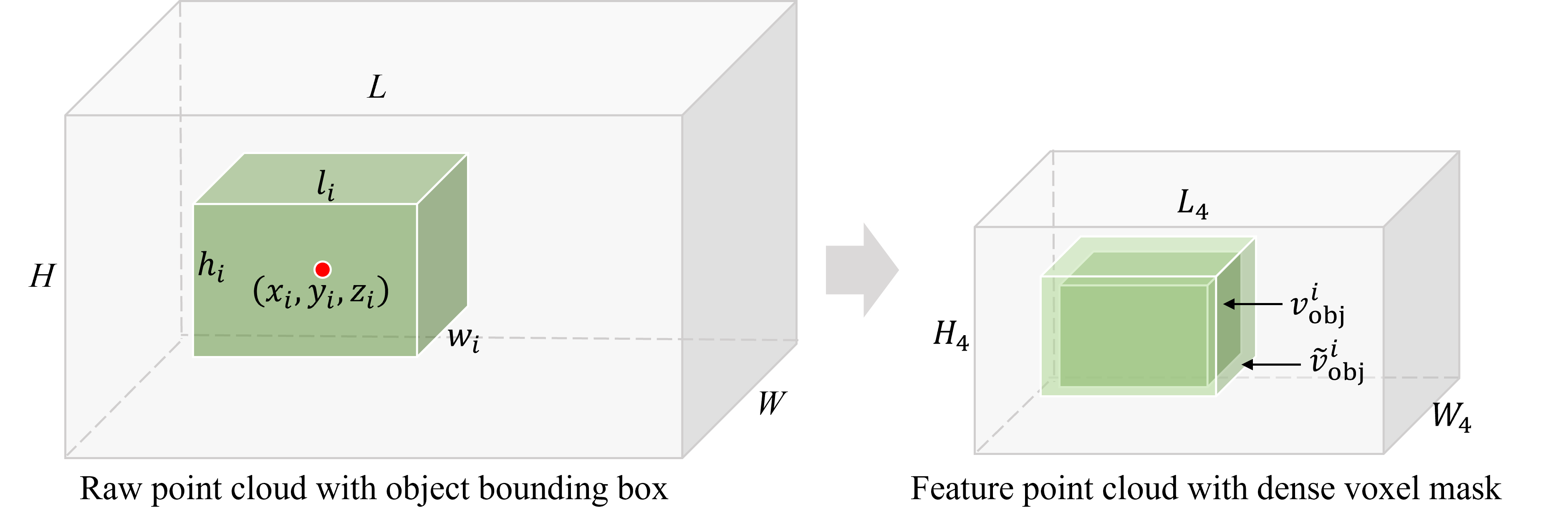}
    \caption{Dense point mask generation.}
    \label{Fig.5}
\end{figure}

\textit{a) Ground Truth Generation of Voxel-Wise Binary Classification}: The ground truth for voxel-wise binary classification is generated based on the ground truth positions of the object bounding boxes. As illustrated in Fig. \ref{Fig.5}, the object-relevant region of $\textbf{F}_4$ can be preliminarily estimated using the bounding box positions together with the spatial correspondence between the original point cloud and $\textbf{F}_4$. Let the spatial dimensions of the entire scene be $L, W, H$, and denote the $i$-th object bounding box by its center $(x_i, y_i, z_i)$ with length $l_i$, width $w_i$, and height $h_i$. The preliminary object-relevant voxels for the $i$-th object in $\textbf{F}_4$ can be represented as
\begin{equation}
\begin{aligned}
v_{\text{obj}}^i = \Big\{ (u,v,w) \,\Big|\, 
& \dfrac{u}{L_4} \in \Big[ \dfrac{x_i}{L} - \dfrac{l_i}{2L},\; \dfrac{x_i}{L} + \dfrac{l_i}{2L} \Big], \\
& \dfrac{v}{W_4} \in \Big[ \dfrac{y_i}{W} - \dfrac{w_i}{2W},\; \dfrac{y_i}{W} + \dfrac{w_i}{2W} \Big], \\
& \dfrac{w}{H_4} \in \Big[ \dfrac{z_i}{H} - \dfrac{h_i}{2H},\; \dfrac{z_i}{H} + \dfrac{h_i}{2H} \Big] 
\Big\},
\end{aligned}
\end{equation}
where $(u,v,w)$ denotes the voxel coordinates in $\textbf{F}_4$, and $(L_4, W_4, H_4)$ represent the spatial dimensions of $\textbf{F}_4$.  

In VirConv-L \cite{ref24}, the feature extractor consists of four stages, each containing two $3\times3\times3$ sparse convolutional layers. Since each convolution enlarges the receptive field by two voxels along each axis, one stage contributes an expansion of four voxels. Consequently, the entire feature extractor enlarges the receptive field by 16 voxels in each direction. Motivated by this observation, we expand the preliminary object-relevant regions accordingly to obtain a more comprehensive representation. The final set of object-relevant voxels for the $i$-th object in $\textbf{F}_4$ is therefore defined as
\begin{equation}
\scalebox{0.9}{$
\begin{aligned}
\tilde{v}_{\text{obj}}^i = \Big\{ (u,v,w) \,\Big|\, 
& \dfrac{u}{L_4} \in \Big[ \dfrac{x_i}{L} - \dfrac{l_i}{2L} - \dfrac{16}{L_4},\; \dfrac{x_i}{L} + \dfrac{l_i}{2L} + \dfrac{16}{L_4} \Big], \\
& \dfrac{v}{W_4} \in \Big[ \dfrac{y_i}{W} - \dfrac{w_i}{2W} - \dfrac{16}{W_4},\; \dfrac{y_i}{W} + \dfrac{w_i}{2W} + \dfrac{16}{W_4} \Big], \\
& \dfrac{w}{H_4} \in \Big[ \dfrac{z_i}{H} - \dfrac{h_i}{2H} - \dfrac{16}{H_4},\; \dfrac{z_i}{H} + \dfrac{h_i}{2H} + \dfrac{16}{H_4} \Big] 
\Big\}.
\end{aligned}
$}
\end{equation}

Finally, the complete ground truth set of object-relevant voxels is obtained by merging all object bounding boxes:
\begin{equation}
v_{\text{obj}} = \bigcup_{i=1}^{N} \tilde{v}_{\text{obj}}^i.
\end{equation}

\textit{b) Architecture of Spatial Compaction Block}:  
As illustrated in Fig. \ref{Fig.4}, the spatial compaction block uses a U-Net-style encoder-decoder architecture with skip connections. It comprises three downsampling units, a bottleneck unit, three upsampling units, and a fusion layer. 

Each downsampling unit consists of three sparse convolution layers $\mathcal{C}$, two batch normalization layers $\mathcal{BN}$, and two ReLU activation functions $\delta_R$. The forward process of a downsampling unit can be expressed as
\begin{equation}
\scalebox{0.9}{$
\textbf{F}_4^{C(l+1)} = \mathcal{C} \Big( \delta_R \big( \mathcal{BN} ( \mathcal{C} ( \delta_R ( \mathcal{BN} ( \mathcal{C} ( F_4^{Cl} ) ) ) ) ) \big) \Big), \quad l \in \{1,2,3\},
$}
\end{equation}
where $\textbf{F}_4^{C1} \equiv \textbf{F}_4^C$. In each unit, the first convolution layer expands the number of channels, the second enhances feature representation, and the third applies spatial downsampling. The kernel sizes of these layers are set to $3\times3\times3, 3\times3\times3,$ and $ 2\times2\times2$, respectively, with corresponding strides of $1,1,$ and $2$. The output channels of the downsampling units are set to $16,32,$ and $64$ respectively.  

The bottleneck unit bridges the downsampling path to the upsampling path, with forward pass
\begin{equation}
\textbf{F}_4^{C5} = \delta_R \big( \mathcal{BN} ( \mathcal{C} ( \delta_R ( \mathcal{BN} ( \mathcal{C} ( \textbf{F}_4^{C4} ) ) ) ) ) \big),
\end{equation}
where both convolution layers have 128 output channels, a kernel size of  $3\times3\times3$, and a stride of 1.  

The decoder consists of three upsampling units and the forward pass of an upsampling unit is

\begin{equation}
\begin{aligned}
\textbf{F}_4^{C(l+1)} 
&= \delta_R \Big( \mathcal{BN} \big( \mathcal{C} \big( 
      \delta_R \big( \mathcal{BN} \big( 
      \mathcal{C} \big( \mathcal{C}_T(\textbf{F}_4^{Cl}) \,\|\, \textbf{F}_4^{C(9-l)} \big) 
      \big) \big) \big) \big) \Big), \\
& \quad l \in \{5,6,7\}.
\end{aligned}
\end{equation}
where `$\|$' denotes channel-wise concatenation. The transposed sparse convolution layer $\mathcal{C}_T$ uses a kernel size of $2\times2\times2$ and a stride of 2. Within each upsampling unit, the first sparse convolution layer reduces the number of channels, while the second enhances feature representation. Both sparse convolution layers use a kernel size of $2\times2\times2$ and a stride of 2. The output channels of the upsampling units are set as $64, 32,$ and $ 16$, respectively.  

Finally, the decoder output, denoted as $\textbf{F}_4^{C8}$, is passed through an additional sparse convolution layer to achieve channel dimensionality reduction
\begin{equation}
\textbf{F}_4^{C9} = \mathcal{C} ( \textbf{F}_4^{C8} ),
\end{equation}
where the sparse convolution has one output channel, a kernel size of $1\times1\times1$, and stride of 1.  

Considering that the feature point cloud is sparse, while the voxel-wise binary classification ground truth is dense, we first compute the intersection between the dense ground truth and the spatial coordinates of $\textbf{F}_4^{C9}$ to generate a sparse ground truth for training. Simultaneously, the attributes of $\textbf{F}_4^{C9}$ are passed through a sigmoid activation function to obtain probabilistic outputs for binary classification. During training, these probabilistic outputs are multiplied by $\textbf{F}_4^C$ to enable trainable soft classification. During the inference, probabilities are binarized using threshold-based spatial compaction: voxels with probabilities greater than 0.5 are retained, others are pruned. To ensure stability when no voxels exceed the threshold, a failsafe mechanism preserves the top 128 highest-confidence voxels, maintaining valid tensor dimensions throughout the network.
\vspace{-0.3cm} 
\subsection{Channel Encoder}

\begin{figure}
    \centering
    \setlength{\abovecaptionskip}{-0.1cm}
    \includegraphics[width=1\linewidth]{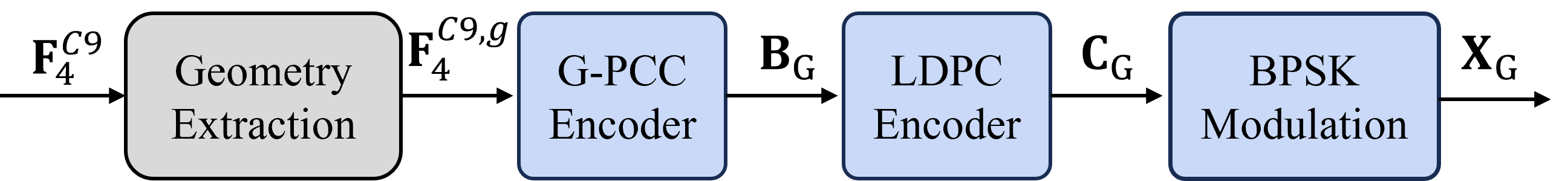}
    \caption{Channel encoding process for geometric information.}
    \label{Fig.6}
\end{figure}

\textbf{Coding for Geometry:} The channel encoding process for geometry is illustrated in Fig.\ref{Fig.6}. 
The geometry information of $\textbf{F}_4^{C9}$, denoted as $\textbf{F}_4^{C9,g}$, is first encoded using a G-PCC encoder to generate the geometry bitstream $\textbf{B}_\text{G}$. 
To ensure reliable error correction, $\textbf{B}_\text{G}$ is then processed by an LDPC encoder, which constructs a parity-check protected sequence with parameters $m=20$, $d_v=2$, and $d_c=5$. 
In this configuration, $m=20$ defines the codeword length factor, $d_v=2$ specifies the variable node degree that determines the number of parity-check connections per variable node in the Tanner graph, and $d_c=5$ represents the check node degree that controls the number of connections per check node and influences the code rate as well as the error correction capability. 
The LDPC-protected geometry bitstream $\textbf{C}_\text{G}$ is finally modulated using Binary Phase Shift Keying (BPSK), mapping each bit to a corresponding carrier phase. 
This conventional digital communication approach ensures strong error resilience and precise reconstruction of 3D structural information at the receiver. Let $\textbf{X}_\text{G}$ denote the modulated signal. After transmission through an additive white Gaussian noise (AWGN) channel, the received signal can be expressed as
\begin{equation}
    \textbf{Y}_\text{G} = \textbf{X}_\text{G} + \eta,
\end{equation}
where $\eta \sim \mathcal{N}\!\left(0, \frac{1}{2 \cdot 10^{\text{SNR}/10}} \right)$ denotes the Gaussian noise term parameterized by the SNR.

\begin{figure*}
    \centering
    \setlength{\abovecaptionskip}{-0.1cm}
    \includegraphics[width=1\linewidth]{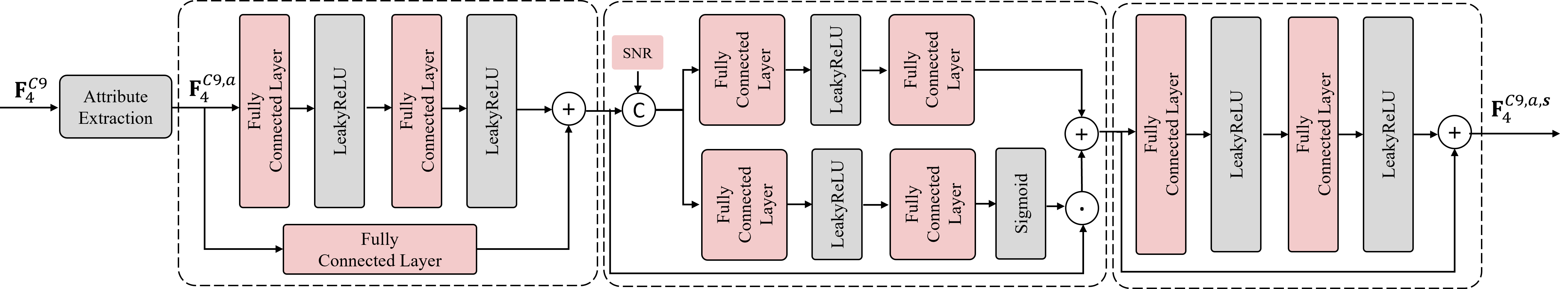}
    \caption{Channel encoding process for attribute information.}
    \label{Fig.7}
\end{figure*}

\textbf{Coding for Attribute:} Unlike the geometric branch, which is encoded using LDPC to provide strong error correction for structured, discrete data and ensure reliable transmission, the attribute is transmitted using DeepJSCC. This approach is better suited for continuous-valued, high-dimensional features, as it jointly optimizes compression and error resilience while adaptively allocating redundancy according to feature importance and channel conditions, achieving more efficient and robust transmission than traditional separate source and channel coding. As shown in Fig.\ref{Fig.7}, the channel encoder for attributes consists of two ResNet blocks and a modulation block. 
The attribute information of $\textbf{F}_4^{C9}$, denoted as $\textbf{F}_4^{C9,a}$, is first fed into the ResNet block, which expands the dimension. 
The forward propagation of this block can be expressed as
\begin{equation}
    \textbf{F}_4^{C9,a,b1} = \mathcal{FC}_{L}\big(\mathcal{FC}_{L}(\textbf{F}_4^{C9,a})\big) + \mathcal{FC}(\textbf{F}_4^{C9,a}),
\end{equation}
where $\textbf{F}_4^{C9,a,b1}$ denotes the output of the first block, and $\mathcal{FC}_{L}$ denotes an FC layer followed by a LeakyReLU activation function. The input and output channel dimensions of the first and second $\mathcal{FC}_{L}$ layers are set to (8, 20) and (20, 40), respectively. The $\mathcal{FC}$ layer has 8 input channels and 40 output channels.

The modulation block then learns to transform these features according to the channel SNR. 
The forward propagation of this block can be expressed as
\begin{equation}
\begin{aligned}
\textbf{F}_4^{C9,a,b2} &= \textbf{F}_4^{C9,a,b1} \cdot \Big(
      \mathcal{FC}_{S}\big(\mathcal{FC}_{L}(\textbf{F}_4^{C9,a,b1} \,\|\, \text{SNR})\big) \\
&\quad + \mathcal{FC}\big(\mathcal{FC}_{L}(\textbf{F}_4^{C9,a,b1} \,\|\, \text{SNR})\big)
      \Big).
\end{aligned}
\end{equation}
where $\textbf{F}_4^{C9,a,b2}$ denotes the output of the second block. All FC layers in this block have 40 output channels. 

Subsequently, a second ResNet block is applied for feature enhancement. 
The forward propagation of this block can be expressed as
\begin{equation}
    \textbf{F}_4^{C9,a,s} = \mathcal{FC}_{L}\big(\mathcal{FC}_{L}(\textbf{F}_4^{C9,a})\big) + \textbf{F}_4^{C9,a},
\end{equation}
where $\textbf{F}_4^{C9,a,s}$ denotes the channel-encoded attribute information to be transmitted. 
All FC layers in this block have 40 output channels.

The channel-encoded attributes are converted into a complex representation for transmission over the wireless channel, where the real part contains the first half of the feature points and the imaginary part contains the second half. 
This complex representation doubles the data rate within the same bandwidth and naturally supports quadrature-domain transmission. 
The complex signal is power-normalized to satisfy the average transmit power constraints and is then subjected to additive complex Gaussian noise, with both real and imaginary components drawn from $\mathcal{N}\!\left(0, \frac{1}{2 \cdot 10^{\text{SNR}/10}}\right)$. Finally, a power denormalization module restores the signal to its original dynamic range. 
This approach provides robustness to channel impairments without relying on digital modulation or error correction codes. 
Together, these two complementary paths provide robust, bandwidth-efficient transmission of both geometric and attribute information, ensuring high-fidelity reconstruction for downstream 3D perception tasks.
\vspace{-0.3cm} 
\subsection{Channel Decoder}

\begin{figure}
    \centering
    \setlength{\abovecaptionskip}{-0.1cm}
    \includegraphics[width=0.8\linewidth]{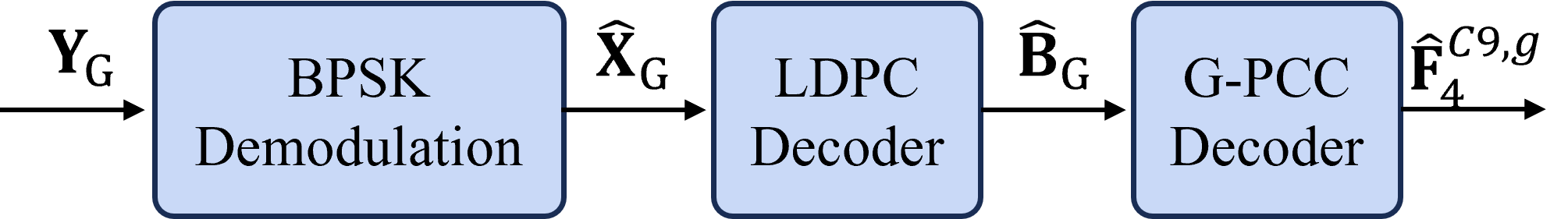}
    \caption{Channel decoding process for geometric information.}
    \label{Fig.8}
\end{figure}

\textbf{Geometry Decoder:} The channel decoding process for geometric information is illustrated in Fig.\ref{Fig.8}. 
At the receiver side, the received signal $\textbf{Y}_\text{G}$ is first demodulated to obtain the LDPC-protected bitstream $\hat{\textbf{X}}_\text{G}$. 
This bitstream is then passed through the LDPC decoder for error correction, yielding the corrected binary bitstream $\hat{\textbf{B}}_\text{G}$. $\hat{\textbf{B}}_\text{G}$ consists of binary values representing the transmitted geometric information in encoded form. It is subsequently delivered to a G-PCC decoder, which reconstructs the original syntax elements of the geometric information for further processing.

\begin{figure*}
    \centering
    \setlength{\abovecaptionskip}{-0.1cm}
    \includegraphics[width=1\linewidth]{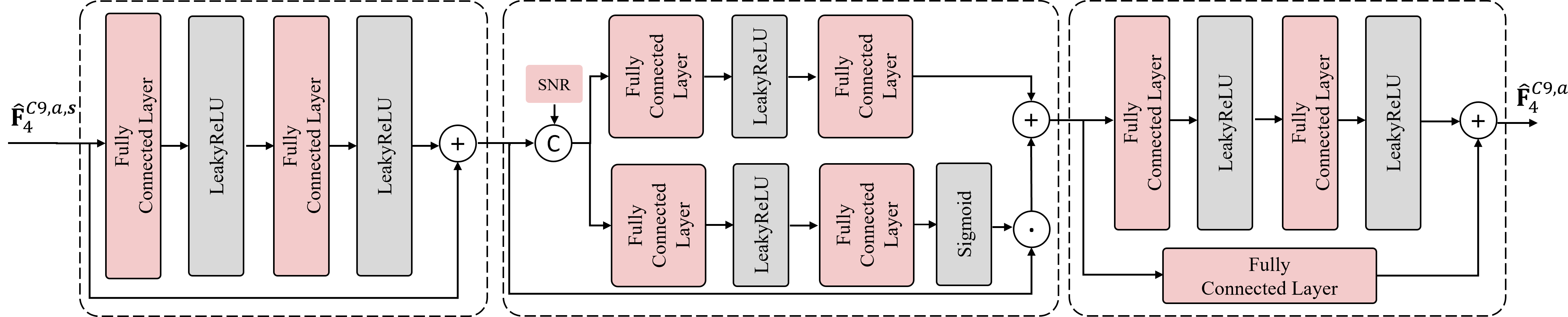}
    \caption{Channel decoding process for attribute information.}
    \label{Fig.9}
\end{figure*}

\textbf{Attribute Decoder:}  Unlike the geometric information decoded from a binary LDPC bitstream, the attribute information is directly reconstructed after channel decoding as a compact feature representation. This compact representation retains the essential information of the original attribute source for further processing. As illustrated in Fig.\ref{Fig.9}, the channel decoder for attribute information is designed symmetrically to its encoder. 
The first ResNet block suppresses noise and refines the received features. 
The forward propagation of this block can be expressed as
\begin{equation}
    \hat{\textbf{F}}_4^{C9,a,b1} = \mathcal{FC}_L \big( \mathcal{FC}_L(\hat{\textbf{F}}_4^{C9,a,s}) \big) + \hat{\textbf{F}}_4^{C9,a,s},
\end{equation}
where $\hat{\textbf{F}}_4^{C9,a,s}$ denotes the received attribute information and $\hat{\textbf{F}}_4^{C9,a,b1}$ denotes the output of the first block. 
All FC layers in this block have 40 output channels. Guided by the channel condition (i.e., SNR), the modulation block applies the inverse transformation of its encoding counterpart, which can be formulated as
\begin{equation}
\begin{aligned}
\hat{\textbf{F}}_4^{C9,a,b2} &= \hat{\textbf{F}}_4^{C9,a,b1} \cdot \Big(
      \mathcal{FC}_S\big(\mathcal{FC}_L(\hat{\textbf{F}}_4^{C9,a,b1} \,\|\, \text{SNR})\big) \\
&\quad + \mathcal{FC}\big(\mathcal{FC}_L(\hat{\textbf{F}}_4^{C9,a,b1} \,\|\, \text{SNR})\big)
      \Big).
\end{aligned}
\end{equation}
where all FC layers in this block produce 40 output channels. Finally, the second ResNet block reverses the encoder’s expansion process by reducing the feature dimensionality to its original size, which removes the redundancy that was introduced for protection. The forward propagation of this block can be formulated as
\begin{equation}
    \hat{\textbf{F}}_4^{C9,a} = \mathcal{FC}_L \big( \mathcal{FC}_L(\hat{\textbf{F}}_4^{C9,a,b2}) \big) + \mathcal{FC}(\hat{\textbf{F}}_4^{C9,a,b2}),
\end{equation}
where the input and output channel dimensions of the first and second $\mathcal{FC}_L$ layers are set to (40, 20) and (20, 8), respectively. The $\mathcal{FC}$ layer has 40 input channels and 8 output channels.
\vspace{-0.3cm} 
\subsection{Source Decoder}
\begin{figure}
    \centering
    \setlength{\abovecaptionskip}{-0.1cm}
    \includegraphics[width=1\linewidth]{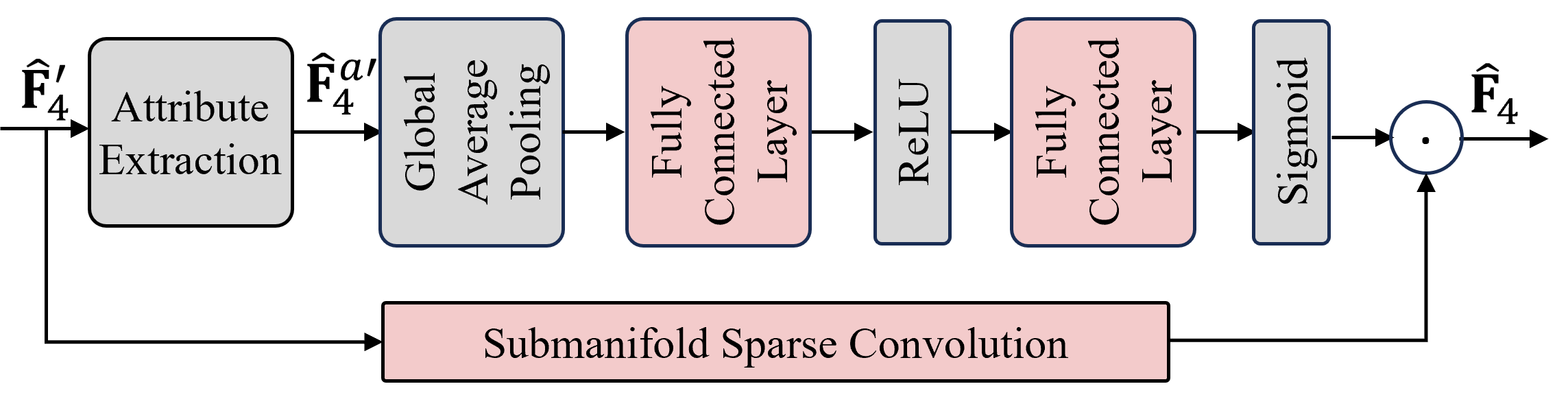}
    \caption{Channel decompaction block.}
    \label{Fig.10}
\end{figure}
Unlike the source encoder, the source decoder focuses solely on reconstructing the channels and recovering the feature representation compatible with subsequent detection steps. As shown in Fig.\ref{Fig.10}, $\hat{\textbf{F}}_4'$ is processed by two parallel branches: the upper branch generates channel-wise weights for feature enhancement, while the lower branch expands the number of channels. In the upper branch, the attribute information is first extracted from $\hat{\textbf{F}}_4'$, resulting in $\hat{\textbf{F}}_4^{a'} \in \mathbb{R}^{n \times 8}$. $\hat{\textbf{F}}_4^{a'}$ is then passed through a GAP layer to produce channel-wise descriptors. Next, two FC layers followed by ReLU and sigmoid activation functions, i.e., $\mathcal{FC}_R$ and $\mathcal{FC}_S$, are applied to generate channel weights:
\begin{equation}
\textbf{W}_C' = \mathcal{FC}_S \big( \mathcal{FC}_R ( \mathcal{GAP}(\hat{\textbf{F}}_4^{a'})) \big),
\end{equation}
where $\textbf{W}_C' \in \mathbb{R}^{1 \times 64}$ denotes the generated weight vector. The output dimensions of $\mathcal{FC}_R$ and $\mathcal{FC}_S$ are set to 48 and 64, respectively. 

In the lower branch of Fig.\ref{Fig.10}, $\hat{\textbf{F}}_4'$ is fed into a submanifold sparse convolution layer $\mathcal{C}_S$ to expand the number of channels. Finally, the generated weight vector $\textbf{W}_C'$ is applied to the attribute channels of the expanded feature to update them, yielding the reconstructed feature $\hat{\textbf{F}}_4$, which can be formulated as
\begin{equation}
\hat{\textbf{F}}_4 = \textbf{W}_C' \cdot \mathcal{C}_S (\hat{\textbf{F}}_4').
\end{equation}
\vspace{-0.3cm} 
\subsection{Feature Point Cloud Upsampling}

\begin{figure*}
    \centering
    \setlength{\abovecaptionskip}{-0.1cm}
    \includegraphics[width=1\linewidth]{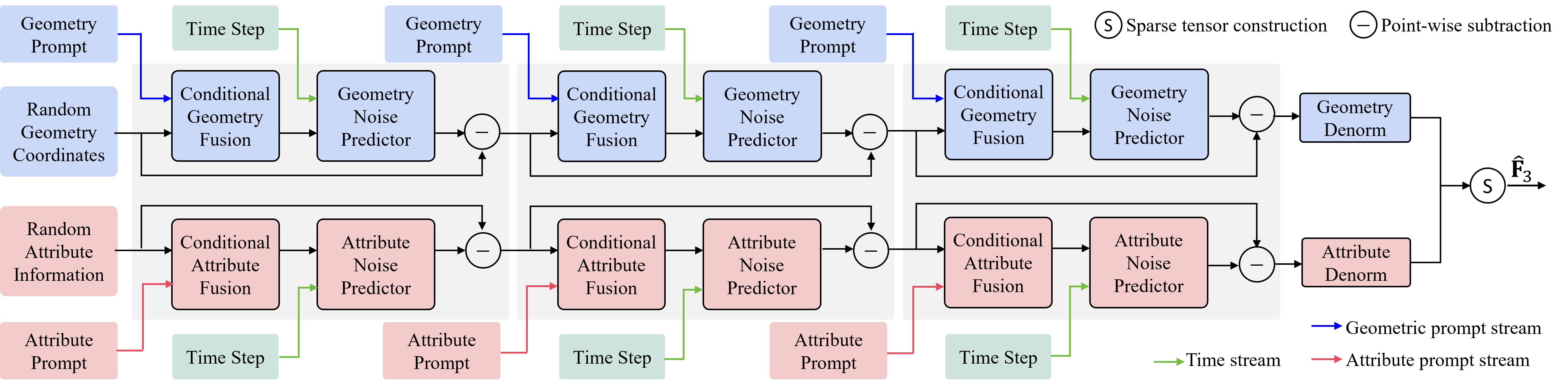}
    \caption{Illustration of diffusion model-based feature point cloud upsampling module.}
    \label{Fig.11}
\end{figure*}

The objective of feature point cloud upsampling is to generate a high-resolution feature point cloud $\hat{\textbf{F}}_3$ from the low-resolution feature point cloud $\hat{\textbf{F}}_4$. The overall pipeline of the proposed diffusion model-based feature point cloud upsampling module is shown in Fig. \ref{Fig.11}. 

\begin{figure}
    \centering
    \setlength{\abovecaptionskip}{-0.1cm}
    \includegraphics[width=1\linewidth]{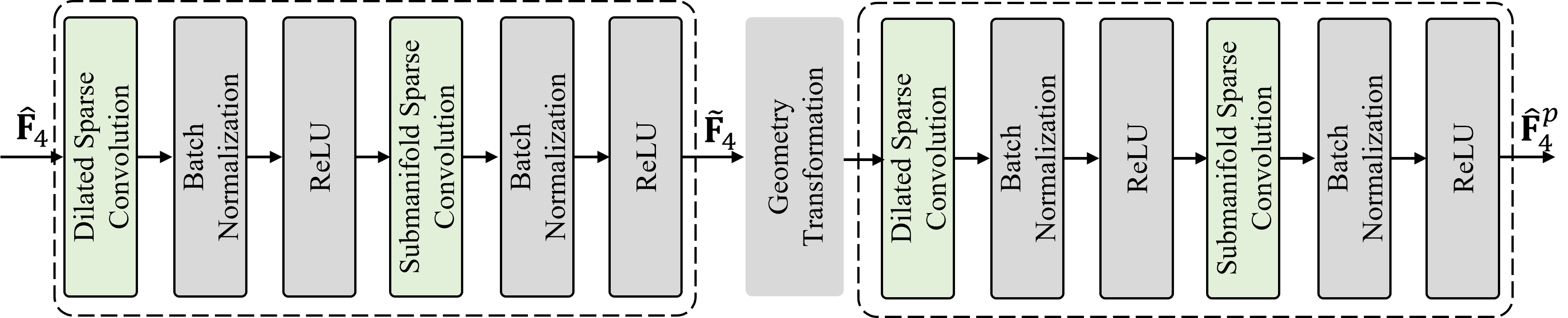}
    \caption{Illustration of prompt generation process.}
    \label{Fig.12}
\end{figure}

\textbf{Prompt Generation:} $\hat{\textbf{F}}_4$ is first sent to the prompt generation block to obtain the enhanced feature $\hat{\textbf{F}}_4^p$. As illustrated in Fig.\ref{Fig.12}, this process can be expressed as
\begin{equation}
    \tilde{\textbf{F}}_4 = \delta_R \big( \mathcal{BN} \big( \mathcal{C}_S( \delta_R( \mathcal{BN}( \mathcal{C}_D(\hat{\textbf{F}}_4) )) ) \big) \big),
\end{equation}
where the dilated sparse convolution layer $\mathcal{C}_D$ is configured with 64 output channels, a kernel size of $3\times 3\times 3$, a stride of 1, and a dilation rate of 2. The $\mathcal{C}_S$ layer is configured with 64 output channels, a kernel size of $3\times 3\times 3$, and a stride of 1. Compared with $\hat{\textbf{F}}_4$, the enhanced feature $\tilde{\textbf{F}}_4$ contains more nonempty voxels, and is therefore used in the following voxel ROI pooling operation to improve the learning capability of the cloud device.

The feature $\tilde{\textbf{F}}_4$ is then fed into a geometry transformation layer $\mathcal{T}$, which expands the geometry coordinates twice to match the spatial resolution of $\textbf{F}_3$. The transformed feature is further processed by another set of dilated and submanifold sparse convolution layers, together with BN and ReLU, to generate the final enhanced feature $\hat{\textbf{F}}_4^p$:
\begin{equation}
    \hat{\textbf{F}}_4^p = \delta_R \big( \mathcal{BN} ( \mathcal{C}_S ( \delta_R ( \mathcal{BN} ( \mathcal{C}_D ( \mathcal{T}(\tilde{\textbf{F}}_4) )))))) \big),
\end{equation}
where the $\mathcal{C}_D$ layer is configured with 64 output channels, a kernel size of $3\times 3\times 3$, a stride of 1, and a dilation rate of 2. The $\mathcal{C}_S$ layer is configured with 64 output channels, a kernel size of $3\times 3\times 3$, and a stride of 1.

\textbf{Conditional Fusion:} Next, $\hat{\textbf{F}}_4^p$ is decomposed into its geometry prompt $\hat{\textbf{F}}_4^{p,g}$ and attribute prompt $\hat{\textbf{F}}_4^{p,a}$, which are then normalized to obtain $\hat{\textbf{F}}_4^{p,g_{\text{norm}}}$ and $\hat{\textbf{F}}_4^{p,a_{\text{norm}}}$. These normalized prompts are combined with the noisy geometry $\textbf{N}_m^g$ and noisy attribute $\textbf{N}_m^a$, where $m$ denotes the diffusion timestep, to generate fused geometry and attribute representations. The process can be expressed as
\begin{align}
    \textbf{N}_m^{g_{\text{fuse}}} &= \mathcal{FC}\big( \mathcal{FC}_R(\hat{\textbf{F}}_4^{p,g_{\text{norm}}} \,\|\, \textbf{N}_m^g ) \big), \\
    \textbf{N}_m^{a_{\text{fuse}}} &= \mathcal{FC}\big( \mathcal{FC}_R(\hat{\textbf{F}}_4^{p,a_{\text{norm}}} \,\|\, \textbf{N}_m^a ) \big).
\end{align}
where the output channels of the geometry-related and attribute-related FC layers are set to 3 and 64, respectively.

\textbf{Noise Prediction:} The fused geometry and attribute, together with the temporal embedding, are fed into a geometry and attribute noise predictor to estimate the noise. Specifically, the time embedding generation unit contains two FC layers and a sigmoid Linear Unit (SiLU):
\begin{equation}
    \textbf{T}_{\text{embed}} = \mathcal{FC}\big( \mathcal{FC}_{\text{Si}}(t) \big),
\end{equation}
where $\mathcal{FC}_{\text{Si}}$ denotes an FC layer followed by a SiLU activation function. The geometry/attribute noise predictor consists of three submanifold sparse convolution layers, with the first two followed by a ReLU activation. To meet the input requirements of sparse convolution, we first construct sparse tensors. The combinations $\textbf{T}_{\text{embed}} + \textbf{N}_m^{g_{\text{fuse}}}$ and $\textbf{T}_{\text{embed}} + \textbf{N}_m^{a_{\text{fuse}}}$ form the attribute components of the sparse tensors, while $\textbf{N}_m^{g_{\text{fuse}}}$ serves as the geometry part for both branches. The resulting sparse tensors are denoted as $\textbf{F}_m^{g_{\text{fuse}}}$ and $\textbf{F}_m^{a_{\text{fuse}}}$, respectively. For the geometry and attribute noise prediction branches at the $m$-th diffusion step, the output sparse tensors can be expressed as
\begin{align}
    \tilde{\textbf{N}}_m^g &= \mathcal{C}_S \Big( \delta_R \big( \mathcal{C}_S \big( \delta_R \big( \mathcal{C}_S ( \textbf{F}_m^{g_{\text{fuse}}} ) \big) \big) \big) \Big), \\
    \tilde{\textbf{N}}_m^a &= \mathcal{C}_S \Big( \delta_R \big( \mathcal{C}_S \big( \delta_R \big( \mathcal{C}_S ( \textbf{F}_m^{a_{\text{fuse}}} ) \big) \big) \big) \Big),
\end{align}
where the input and output channel dimensions of the geometry-related convolution layers are set to 3, while those of the attribute-related convolution layers are set to 64. All convolution kernels have a size of $3\times 3\times 3$ with a stride of 1. The predicted noise corresponds to the attribute component of the output sparse tensors.

The predicted noise is subtracted from the noisy input at the current timestep to obtain the denoised output. By iteratively applying this process over $M$ denoising steps, the normalized target point cloud is progressively reconstructed. Finally, denormalization is used to restore the point cloud to its original scale, yielding the final result. In our experiment, $M$ was set to 1.
\vspace{-0.3cm} 
\subsection{Loss Function}
The proposed model involves multiple sub-tasks, including source coding, channel coding, feature point cloud upsampling, and 3D object detection. To ensure stable and efficient training, we use a two-phase strategy. In phase one, we optimize the source coding, feature upsampling, and 3D detection modules using losses for binary mask prediction and detection. Channel compression and feature upsampling are driven by detection performance. In phase two, we train the channel code with a mean squared error (MSE) loss to minimize attribute discrepancies. 

\textbf{Phase One:}  
The phase one loss consists of a binary mask prediction loss $\mathcal{L}_{\text{Bi}}$ and a 3D object detection loss $\mathcal{L}_{\text{Det}}$, expressed as
\begin{equation}
    \mathcal{L}_{\text{S1}} = \mathcal{L}_{\text{Bi}} + \alpha \cdot \mathcal{L}_{\text{Det}},
\end{equation}
where $\alpha$ balances the contribution of $\mathcal{L}_{\text{Det}}$. To address the class imbalance between foreground and background voxels, $\mathcal{L}_{\text{Bi}}$ is formulated using the Focal loss \cite{ref16}:
\begin{equation}
    \mathcal{L}_{\text{Bi}} = - \sum_i \Big[ (1-p_i)^\gamma y_i \log(p_i) + p_i^\gamma (1-y_i)\log(1-p_i) \Big],
\end{equation}
where $y_i \in \{0,1\}$ denotes the ground truth label of the $i$-th voxel, $p_i$ is the predicted probability, and $\gamma$ is the focusing parameter that reduces the relative loss for well-classified examples. In our experiment, $\gamma$ was set to 2.

Following \cite{ref24}, $\mathcal{L}_{\text{Det}}$ consists of two components: the RPN loss and the detection head loss. The RPN loss combines the Focal loss \cite{ref16} for classification with the Huber loss for bounding box regression. Similarly, the detection head loss consists of two parts: the Huber loss for bounding box regression and the Binary Cross Entropy loss for classification.

\textbf{Phase Two:}  
In phase two, the parameters of the attribute-related channel code are optimized using the MSE loss, defined as
\begin{equation}
    \mathcal{L}_{\text{S2}} = \text{MSE}\big( \textbf{Attr}_{\text{transmit}}, \textbf{Attr}_{\text{receive}} \big),
\end{equation}
where $\text{Attr}_{\text{transmit}}$ and $\text{Attr}_{\text{receive}}$ denote the transmitted and received attribute representations at the edge and cloud devices, respectively.

\section{Experimental Results}
\subsection{Datasets and Experimental Settings}

\textbf{Datasets:} Following \cite{ref24}, we evaluate the proposed method on the KITTI dataset \cite{ref51}. The dataset contains 3712 training samples and 3769 validation samples, where each sample consists of an image and its corresponding point cloud. According to the official evaluation protocol, objects are divided into three difficulty levels: easy, moderate, and hard. 

\textbf{Evaluation criteria:} Detection accuracy was measured using the Average Precision (AP) metric under the constraint that recall does not exceed 40\%. An Intersection over Union (IoU) threshold of 0.7 was applied. Both Car BEV AP and Car 3D AP are reported to assess detection performance in BEV and full 3D space, respectively. In addition, we introduce the compression rate (CR) metric to evaluate compression performance, defined as the ratio between the dimensionality of the original feature point cloud and the dimensionality of the compressed representation.

\textbf{Implementation details:} For the object detection task, the voxel size was set to $0.1\,$m, $0.1\,$m, and $0.15\,$m along the $X$, $Y$, and $Z$ axes, respectively. The detection model was trained on an NVIDIA\textsuperscript{\textregistered} GeForce RTX\texttrademark~4090 GPU with an ADAM optimizer. Training used a cyclic learning rate schedule with an initial learning rate of 0.01 for 50 epochs. For optimizing the attribute-related channel code, the model was trained for 20 epochs with ADAM, a fixed learning rate of $1\times10^{-4}$, and weight decay of $1\times10^{-5}$, where the $L_2$ regularization helps mitigate overfitting by penalizing large parameter values. 
\vspace{-0.3cm} 
\subsection{Overall Performance of the Proposed Method}
\begin{table*}[]
\setlength{\abovecaptionskip}{-0.1cm}
\caption{OVERALL PERFORMANCE OF THE PROPOSED METHOD. THE BEST RESULTS ARE SHOWN IN BOLD.}
\resizebox{\textwidth}{!}{
\label{Table I}
\begin{tabular}{c|c|c|c|c|ccc|ccc}
\Xhline{1.0pt} 
\multirow{2}{*}{Scheme} & \multirow{2}{*}{\begin{tabular}[c]{@{}c@{}}Method\end{tabular}} & \multirow{2}{*}{\begin{tabular}[c]{@{}c@{}}Channel\\ SNR (dB)\end{tabular}} & \multirow{2}{*}{\begin{tabular}[c]{@{}c@{}}Data\\ Volume (MB)\end{tabular}} & \multirow{2}{*}{\begin{tabular}[c]{@{}c@{}}CR\end{tabular}} & \multicolumn{3}{c|}{Car BEV AP (R40)}                                                            & \multicolumn{3}{c}{Car 3D AP (R40)}                                                              \\ \cline{6-11} 
                        &                                                                               &                                                                             &                                                                             &                                                                              & \multicolumn{1}{c|}{Easy}             & \multicolumn{1}{c|}{Moderate}         & Hard             & \multicolumn{1}{c|}{Easy}             & \multicolumn{1}{c|}{Moderate}         & Hard             \\ \Xhline{1.0pt} %
1                       & VirConv-L                                                                     & ——                                                                          & 58921                                                                       & 1                                                                            & \multicolumn{1}{c|}{96.08}          & \multicolumn{1}{c|}{92.04}          & 91.44          & \multicolumn{1}{c|}{92.89}          & \multicolumn{1}{c|}{88.07}          & 85.49          \\
2                       & Proposed (VirConv-L + Feature Enhancement)                                               & ——                                                                          & ——                                                                          & ——                                                                           & \multicolumn{1}{c|}{\textbf{97.22}} & \multicolumn{1}{c|}{\textbf{97.13}} & \textbf{92.14} & \multicolumn{1}{c|}{\textbf{96.95}} & \multicolumn{1}{c|}{\textbf{96.60}} & \textbf{91.66} \\
3                       & Proposed (VirConv-L + Source Encoder)                                                    & ——                                                                          & 373.1624                                                                    & 158                                                                          & \multicolumn{1}{c|}{96.95}          & \multicolumn{1}{c|}{96.29}          & 88.76          & \multicolumn{1}{c|}{96.19}          & \multicolumn{1}{c|}{92.62}          & 85.13          \\
4                       & Proposed (VirConv-L + Source Encoder + Geo\_G-PCC  + Attri\_CABAC)                                     & ——                                                                          & 6.5004                                                                      & \textbf{9064}                                                                & \multicolumn{1}{c|}{96.84}          & \multicolumn{1}{c|}{96.22}          & 88.71          & \multicolumn{1}{c|}{96.01}          & \multicolumn{1}{c|}{92.61}          & 85.13          \\ \hline
5                       & T-FFC {[}5{]}                                                                 & ——                                                                          & 11.9046                                                                     & 4949                                                                         & \multicolumn{1}{c|}{96.43}          & \multicolumn{1}{c|}{92.21}          & 89.45          & \multicolumn{1}{c|}{92.84}          & \multicolumn{1}{c|}{85.59}          & 82.59          \\
6                       & A-FFC {[}5{]}                                                                 & ——                                                                          & 80.3980                                                                     & 732                                                                          & \multicolumn{1}{c|}{96.19}          & \multicolumn{1}{c|}{92.03}          & 89.51          & \multicolumn{1}{c|}{92.93}          & \multicolumn{1}{c|}{88.18}          & 85.54          \\ \hline
7                       & Proposed (Overall)                                                                      & 20                                                                          & 341.8359                                                                    & 172                                                                          & \multicolumn{1}{c|}{96.90}          & \multicolumn{1}{c|}{96.24}          & 88.69          & \multicolumn{1}{c|}{96.10}          & \multicolumn{1}{c|}{92.54}          & 85.06          \\
8                       & Proposed (Overall)                                                                      & 15                                                                          & 341.8359                                                                    & 172                                                                          & \multicolumn{1}{c|}{96.82}          & \multicolumn{1}{c|}{96.21}          & 88.68          & \multicolumn{1}{c|}{96.11}          & \multicolumn{1}{c|}{92.52}          & 85.02          \\
9                       & Proposed (Overall)                                                                      & 10                                                                          & 341.8359                                                                    & 172                                                                          & \multicolumn{1}{c|}{96.89}          & \multicolumn{1}{c|}{96.22}          & 88.69          & \multicolumn{1}{c|}{96.03}          & \multicolumn{1}{c|}{90.29}          & 84.91          \\
10                      & Proposed (Overall)                                                                      & 5                                                                           & 341.8359                                                                    & 172                                                                          & \multicolumn{1}{c|}{96.83}          & \multicolumn{1}{c|}{93.92}          & 88.62          & \multicolumn{1}{c|}{95.91}          & \multicolumn{1}{c|}{90.13}          & 82.55          \\
11                      & Proposed (Overall)                                                                      & 0                                                                           & 341.8359                                                                    & 172                                                                          & \multicolumn{1}{c|}{96.61}          & \multicolumn{1}{c|}{91.24}          & 83.66          & \multicolumn{1}{c|}{93.17}          & \multicolumn{1}{c|}{86.96}          & 77.25          \\ \Xhline{1.0pt} %
\end{tabular}
}
\end{table*}

The performance of the proposed method is summarized in Table \ref{Table I}, which highlights the following advantages. a) Improved detection accuracy: Compared with the original detection model (Scheme-1), VirConv-L equipped with the proposed feature enhancement (Scheme-2) achieved 3D AP improvements of 4.06\%, 8.53\%, and 6.17\% for easy, moderate, and hard objects, respectively. b) Cloud-edge adaptability with feature upsampling: To better support collaborative object detection in cloud-edge scenarios, we introduced a feature point cloud upsampling module that generates a high-resolution feature $\hat{\textbf{F}}_3$ from the reconstructed low-resolution feature $\hat{\textbf{F}}_4$. Compared to directly using the ground-truth $\textbf{F}_3$ (Scheme-2), the upsampled $\hat{\textbf{F}}_3$ (Scheme-3) substantially reduced data transmission overhead while still surpassing the baseline detection performance of the original model (Scheme-1). c) Compatibility with entropy coding: Although no dedicated entropy coding method was designed in this work due to the inclusion of channel coding, the source-encoded features remain in point cloud format and can thus be directly processed by existing entropy coding methods (Scheme-4). In particular, the geometry component of the point cloud is compressed using the G-PCC octree-based coding method \cite{ref8}, while the attribute component is encoded with Context-Based Adaptive Binary Arithmetic Coding (CABAC) \cite{ref52}. Experimental results show that the integration of these two coding techniques enables the proposed framework to achieve a compression rate of 9064×, which substantially outperforms our previous work \cite{ref5} (Scheme-5 and Scheme-6). d) Enhanced transmission reliability: Finally, the channel code (Scheme-7 to Scheme-11) improved robustness by introducing structured redundancy. Results demonstrate that even at an SNR of 0 dB, the proposed system maintained a detection performance of no less than 86.96\% on the moderate object while achieving a compression rate of 172×.
\vspace{-0.3cm} 
\subsection{Performance with Different Parameter Settings}

\begin{table}[]
\setlength{\abovecaptionskip}{-0.1cm}
\caption{PERFORMANCE WITH DIFFERENT MASK LABELS. THE BEST RESULTS ARE SHOWN IN BOLD.}
\resizebox{\columnwidth}{!}{
\label{Table 2}
\begin{tabular}{c|c|ccc|ccc}
\Xhline{1.0pt} %
\multirow{2}{*}{Scheme} & \multirow{2}{*}{\begin{tabular}[c]{@{}c@{}}Expansion\\ Size\end{tabular}} & \multicolumn{3}{c|}{Car BEV AP (R40)}                                                            & \multicolumn{3}{c}{Car 3D AP (R40)}                                                              \\ \cline{3-8} 
                        &                                                                           & \multicolumn{1}{c|}{Easy}             & \multicolumn{1}{c|}{Moderate}         & Hard             & \multicolumn{1}{c|}{Easy}             & \multicolumn{1}{c|}{Moderate}         & Hard             \\ \Xhline{1.0pt} %
1                       & ——                                                                        & \multicolumn{1}{c|}{96.08}          & \multicolumn{1}{c|}{92.04}          & \textbf{91.44} & \multicolumn{1}{c|}{92.89}          & \multicolumn{1}{c|}{88.07}          & 85.49          \\
2                       & 0                                                                         & \multicolumn{1}{c|}{95.97}          & \multicolumn{1}{c|}{\textbf{92.18}} & 87.24          & \multicolumn{1}{c|}{95.00}          & \multicolumn{1}{c|}{86.51}          & 83.44          \\
3                       & 8                                                                         & \multicolumn{1}{c|}{96.23}          & \multicolumn{1}{c|}{92.07}          & 89.50          & \multicolumn{1}{c|}{95.35}          & \multicolumn{1}{c|}{\textbf{88.57}} & 83.93          \\
4                       & 16                                                                        & \multicolumn{1}{c|}{96.22}          & \multicolumn{1}{c|}{92.04}          & 89.50          & \multicolumn{1}{c|}{\textbf{95.44}} & \multicolumn{1}{c|}{88.53}          & \textbf{85.87} \\
5                       & 24                                                                        & \multicolumn{1}{c|}{\textbf{96.25}} & \multicolumn{1}{c|}{92.06}          & 89.49          & \multicolumn{1}{c|}{95.33}          & \multicolumn{1}{c|}{88.43}          & 85.76          \\
6                       & 32                                                                        & \multicolumn{1}{c|}{96.22}          & \multicolumn{1}{c|}{91.98}          & 89.46          & \multicolumn{1}{c|}{93.41}          & \multicolumn{1}{c|}{88.43}          & 85.81          \\ \Xhline{1.0pt} %
\end{tabular}
}
\end{table}

\textbf{Mask labels:} As mentioned above, convolution operations in the feature extraction process tend to enlarge the receptive field of object-related regions. To capture a more complete representation, we apply a region expansion operation to the preliminary object-relevant area. Table \ref{Table 2} summarizes the detection performance under varying expansion sizes. Scheme 1 corresponds to the original VirConv-L detection model, while Schemes 2-6 filter object-relevant voxels with different mask labels during the testing phase. As the expansion size increased, detection performance initially improved and eventually saturated when the expansion size reached 16.

\begin{table}[]
\setlength{\abovecaptionskip}{-0.1cm}
\caption{PERFORMANCE WITH DIFFERENT NUMBER OF CHANNELS. THE BEST RESULTS ARE SHOWN IN BOLD.}
\resizebox{\columnwidth}{!}{ 
\label{Table 3}
\begin{tabular}{c|c|lll|lll}
\Xhline{1.0pt} %
\multirow{2}{*}{Scheme} & \multirow{2}{*}{\begin{tabular}[c]{@{}c@{}}Number of\\ Channels\end{tabular}} & \multicolumn{3}{c|}{Car BEV AP (R40)}                                                            & \multicolumn{3}{c}{Car 3D AP (R40)}                                                                      \\ \cline{3-8} 
                        &                                                                           & \multicolumn{1}{c|}{Easy}    & \multicolumn{1}{c|}{Moderate}         & \multicolumn{1}{c|}{Hard} & \multicolumn{1}{c|}{Easy}             & \multicolumn{1}{c|}{Moderate}         & \multicolumn{1}{c}{Hard} \\ \Xhline{1.0pt} %
1                       & 64                                                                        & \multicolumn{1}{c|}{96.08} & \multicolumn{1}{c|}{92.04}          & \textbf{91.44}          & \multicolumn{1}{l|}{92.89}          & \multicolumn{1}{c|}{\textbf{88.07}}          & \textbf{85.49}                  \\
2                       & 2                                                                         & \multicolumn{1}{c|}{92.80} & \multicolumn{1}{c|}{88.89} & 86.20                   & \multicolumn{1}{c|}{89.30}          & \multicolumn{1}{c|}{82.90}          & 80.11                  \\
3                       & 4                                                                         & \multicolumn{1}{c|}{95.41} & \multicolumn{1}{c|}{91.67}          & 89.18                   & \multicolumn{1}{c|}{92.37}          & \multicolumn{1}{c|}{85.83} & 83.21                  \\
4                       & 6                                                                         & \multicolumn{1}{c|}{96.08} & \multicolumn{1}{c|}{92.05}          & 89.48                   & \multicolumn{1}{c|}{93.08} & \multicolumn{1}{c|}{88.04}          & 85.40        \\
5                       & 8                                                                         & \multicolumn{1}{c|}{\textbf{96.17}} & \multicolumn{1}{c|}{\textbf{92.11}}          & 91.34                   & \multicolumn{1}{c|}{\textbf{93.08}}          & \multicolumn{1}{c|}{87.87}          & 85.25                  \\ \Xhline{1.0pt} %
\end{tabular}
}
\end{table}

\textbf{Number of channels:} The channel compaction process was designed to extract intrinsic representations from high-dimensional features. Table \ref{Table 3} shows the detection performance with varying number of channels. Scheme 1 corresponds to the original VirConv-L detection model, while Schemes 2-5 present the test performance of models re-trained with different channel configurations on the KITTI validation dataset. Scheme 5 achieved near-lossless detection while using the minimum number of channels.

\begin{table}[]
\setlength{\abovecaptionskip}{-0.1cm}
\caption{PERFORMANCE WITH DIFFERENT LOSS WEIGHTS. THE BEST RESULTS ARE SHOWN IN BOLD. }
\resizebox{\columnwidth}{!}{ 
\label{Table 5}
\begin{tabular}{c|c|ccc|ccc}
\Xhline{1.0pt} %
\multirow{2}{*}{Scheme} & \multirow{2}{*}{\begin{tabular}[c]{@{}c@{}}Loss\\ Weight\end{tabular}} & \multicolumn{3}{c|}{Car BEV AP (R40)}                                  & \multicolumn{3}{c}{Car 3D AP (R40)}                                    \\ \cline{3-8} 
                        &                                                                        & \multicolumn{1}{c|}{Easy}    & \multicolumn{1}{c|}{Moderate} & Hard    & \multicolumn{1}{c|}{Easy}    & \multicolumn{1}{c|}{Moderate} & Hard    \\ \Xhline{1.0pt} %
1                       & ——                                                                     & \multicolumn{1}{c|}{96.08} & \multicolumn{1}{c|}{92.04}  & 91.44 & \multicolumn{1}{c|}{92.89} & \multicolumn{1}{c|}{88.07}  & 85.49 \\
2                       & 1                                                                      & \multicolumn{1}{c|}{96.16} & \multicolumn{1}{c|}{95.93}  & 91.34 & \multicolumn{1}{c|}{96.02} & \multicolumn{1}{c|}{95.64}  & 90.70 \\
3                       & 10                                                                     & \multicolumn{1}{c|}{\textbf{97.22}} & \multicolumn{1}{c|}{\textbf{97.13}}  & \textbf{92.14} & \multicolumn{1}{c|}{\textbf{96.95}} & \multicolumn{1}{c|}{\textbf{96.60}}  & \textbf{91.66} \\
4                       & 100                                                                    & \multicolumn{1}{c|}{97.18} & \multicolumn{1}{c|}{96.92}  & 91.87 & \multicolumn{1}{c|}{96.95} & \multicolumn{1}{c|}{94.18}  & 89.11 \\ \Xhline{1.0pt} %
\end{tabular}}
\end{table}

\textbf{Loss weights:} The hyperparameter $\alpha$ controls the trade-off between the object detection loss and the binary classification loss. By setting $\alpha$ to 10, the magnitudes of the two loss terms were balanced. As shown in Table \ref{Table 5}, the model achieved its best performance when the two losses were of comparable scale (Scheme 3). In contrast, assigning disproportionately high weights to either loss term (Schemes 2 and 4) led to performance degradation.

\begin{figure*}
    \centering
    \setlength{\abovecaptionskip}{-0.1cm}
    \includegraphics[width=1\linewidth]{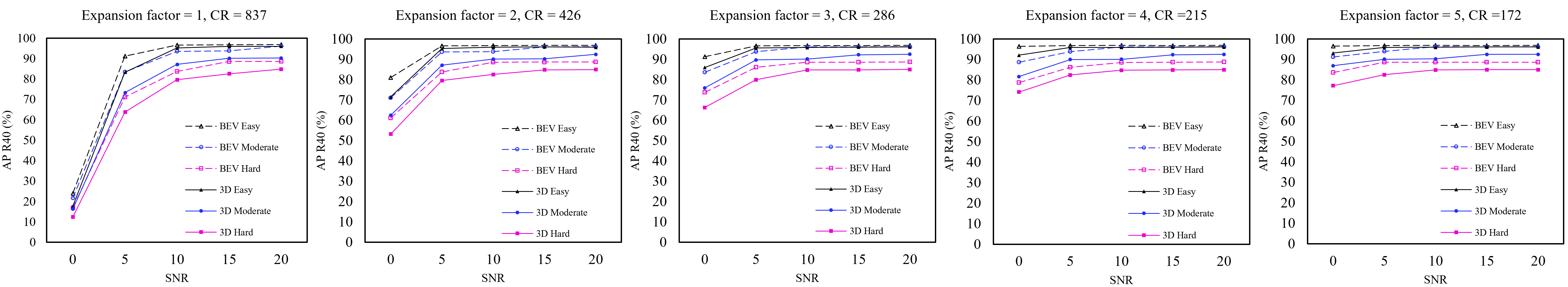}
    \caption{Performance with different attribute expansion factors.}
    \label{Fig.13}
\end{figure*}

\textbf{Expansion factors:} The core principle of the attribute-related channel code is to mitigate channel interference by injecting controlled redundancy at the edge-side encoding stage, which enhances the quality of received signals and improves overall system robustness. Fig \ref{Fig.13} summarizes the detection performance under varying expansion factors, demonstrating the effectiveness of redundancy in strengthening channel reliability.
\vspace{-0.3cm} 
\subsection{Ablation Study}
\begin{table}[]
\setlength{\abovecaptionskip}{-0.1cm}
\caption{ABLATION STUDY OF THE CHANNEL CODE. THE BEST RESULTS ARE SHOWN IN BOLD. “\ding{51}” INDICATES THAT THE CHANNEL CODE IS ENABLED, WHILE “\ding{55}” INDICATES THAT IT IS DISABLED.}
\resizebox{\columnwidth}{!}{ 
\label{Table 7}
\begin{tabular}{c|c|cc|ccc|ccc}
\Xhline{1.0pt} %
\multirow{2}{*}{Scheme} & \multirow{2}{*}{\begin{tabular}[c]{@{}c@{}}Channel\\ SNR (dB)\end{tabular}} & \multicolumn{2}{c|}{Channel Code}        & \multicolumn{3}{c|}{Car BEV AP (R40)}                                                            & \multicolumn{3}{c}{Car 3D AP (R40)}                                                              \\ \cline{3-10} 
                        &                                                                             & \multicolumn{1}{c|}{Geometry} & Attribute & \multicolumn{1}{c|}{Easy}             & \multicolumn{1}{c|}{Moderate}         & Hard             & \multicolumn{1}{c|}{Easy}             & \multicolumn{1}{c|}{Moderate}         & Hard             \\ \Xhline{1.0pt} %
1                       & Noiseless                                                                   & \multicolumn{1}{c|}{\ding{55}}        & \ding{55}         & \multicolumn{1}{c|}{\textbf{96.92}} & \multicolumn{1}{c|}{\textbf{96.26}} & \textbf{91.44} & \multicolumn{1}{c|}{\textbf{96.19}} & \multicolumn{1}{c|}{\textbf{92.55}} & \textbf{85.49} \\ \hline
2                       & 20                                                                          & \multicolumn{1}{c|}{\ding{55}}        & \ding{55}         & \multicolumn{1}{c|}{0.00}           & \multicolumn{1}{c|}{0.00}           & 0.00           & \multicolumn{1}{c|}{0.00}           & \multicolumn{1}{c|}{0.00}           & 0.00           \\
3                       & 15                                                                          & \multicolumn{1}{c|}{\ding{55}}        & \ding{55}         & \multicolumn{1}{c|}{0.00}           & \multicolumn{1}{c|}{0.00}           & 0.00           & \multicolumn{1}{c|}{0.00}           & \multicolumn{1}{c|}{0.00}           & 0.00           \\
4                       & 10                                                                          & \multicolumn{1}{c|}{\ding{55}}        & \ding{55}         & \multicolumn{1}{c|}{0.00}           & \multicolumn{1}{c|}{0.00}           & 0.00           & \multicolumn{1}{c|}{0.00}           & \multicolumn{1}{c|}{0.00}           & 0.00           \\
5                       & 5                                                                           & \multicolumn{1}{c|}{\ding{55}}        & \ding{55}         & \multicolumn{1}{c|}{0.00}           & \multicolumn{1}{c|}{0.00}           & 0.00           & \multicolumn{1}{c|}{0.00}           & \multicolumn{1}{c|}{0.00}           & 0.00           \\
6                       & 0                                                                           & \multicolumn{1}{c|}{\ding{55}}        & \ding{55}         & \multicolumn{1}{c|}{0.00}           & \multicolumn{1}{c|}{0.00}           & 0.00           & \multicolumn{1}{c|}{0.00}           & \multicolumn{1}{c|}{0.00}           & 0.00           \\ \hline
7                       & 20                                                                          & \multicolumn{1}{c|}{\ding{51}}        & \ding{55}         & \multicolumn{1}{c|}{0.00}           & \multicolumn{1}{c|}{0.00}           & 0.00           & \multicolumn{1}{c|}{0.00}           & \multicolumn{1}{c|}{0.00}           & 0.00           \\
8                       & 15                                                                          & \multicolumn{1}{c|}{\ding{51}}        & \ding{55}         & \multicolumn{1}{c|}{0.00}           & \multicolumn{1}{c|}{0.00}           & 0.00           & \multicolumn{1}{c|}{0.00}           & \multicolumn{1}{c|}{0.00}           & 0.00           \\
9                       & 10                                                                          & \multicolumn{1}{c|}{\ding{51}}        & \ding{55}         & \multicolumn{1}{c|}{0.00}           & \multicolumn{1}{c|}{0.00}           & 0.00           & \multicolumn{1}{c|}{0.00}           & \multicolumn{1}{c|}{0.00}           & 0.00           \\
10                      & 5                                                                           & \multicolumn{1}{c|}{\ding{51}}        & \ding{55}         & \multicolumn{1}{c|}{0.00}           & \multicolumn{1}{c|}{0.00}           & 0.00           & \multicolumn{1}{c|}{0.00}           & \multicolumn{1}{c|}{0.00}           & 0.00           \\
11                      & 0                                                                           & \multicolumn{1}{c|}{\ding{51}}        & \ding{55}         & \multicolumn{1}{c|}{0.00}           & \multicolumn{1}{c|}{0.00}           & 0.00           & \multicolumn{1}{c|}{0.00}           & \multicolumn{1}{c|}{0.00}           & 0.00           \\ \hline
12                      & 20                                                                          & \multicolumn{1}{c|}{\ding{55}}        & \ding{51}         & \multicolumn{1}{c|}{0.00}           & \multicolumn{1}{c|}{0.00}           & 0.00           & \multicolumn{1}{c|}{0.00}           & \multicolumn{1}{c|}{0.00}           & 0.06           \\
13                      & 15                                                                          & \multicolumn{1}{c|}{\ding{55}}        & \ding{51}         & \multicolumn{1}{c|}{0.00}           & \multicolumn{1}{c|}{0.00}           & 0.00           & \multicolumn{1}{c|}{0.00}           & \multicolumn{1}{c|}{0.00}           & 0.00           \\
14                      & 10                                                                          & \multicolumn{1}{c|}{\ding{55}}        & \ding{51}         & \multicolumn{1}{c|}{0.00}           & \multicolumn{1}{c|}{0.00}           & 0.00           & \multicolumn{1}{c|}{0.00}           & \multicolumn{1}{c|}{0.00}           & 0.00           \\
15                      & 5                                                                           & \multicolumn{1}{c|}{\ding{55}}        & \ding{51}         & \multicolumn{1}{c|}{0.00}           & \multicolumn{1}{c|}{0.00}           & 0.00           & \multicolumn{1}{c|}{0.00}           & \multicolumn{1}{c|}{0.00}           & 0.00           \\
16                      & 0                                                                           & \multicolumn{1}{c|}{\ding{55}}        & \ding{51}& \multicolumn{1}{c|}{0.00}           & \multicolumn{1}{c|}{0.00}           & 0.00           & \multicolumn{1}{c|}{0.00}           & \multicolumn{1}{c|}{0.00}           & 0.00           \\ \hline
17                      & 20                                                                          & \multicolumn{1}{c|}{\ding{51}}        & \ding{51}         & \multicolumn{1}{c|}{96.90}          & \multicolumn{1}{c|}{96.24}          & 88.69          & \multicolumn{1}{c|}{96.10}          & \multicolumn{1}{c|}{92.54}          & 85.06          \\
18                      & 15                                                                          & \multicolumn{1}{c|}{\ding{51}}        & \ding{51}         & \multicolumn{1}{c|}{96.82}          & \multicolumn{1}{c|}{96.21}          & 88.68          & \multicolumn{1}{c|}{96.11}          & \multicolumn{1}{c|}{92.52}          & 85.02          \\
19                      & 10                                                                          & \multicolumn{1}{c|}{\ding{51}}        & \ding{51}         & \multicolumn{1}{c|}{96.89}          & \multicolumn{1}{c|}{96.22}          & 88.69          & \multicolumn{1}{c|}{96.03}          & \multicolumn{1}{c|}{90.29}          & 84.91          \\
20                      & 5                                                                           & \multicolumn{1}{c|}{\ding{51}}        & \ding{51}         & \multicolumn{1}{c|}{96.83}          & \multicolumn{1}{c|}{93.92}          & 88.62          & \multicolumn{1}{c|}{95.91}          & \multicolumn{1}{c|}{90.13}          & 82.55          \\
21                      & 0                                                                           & \multicolumn{1}{c|}{\ding{51}}        & \ding{51}         & \multicolumn{1}{c|}{96.61}          & \multicolumn{1}{c|}{91.24}          & 83.66          & \multicolumn{1}{c|}{93.17}          & \multicolumn{1}{c|}{86.96}          & 77.25          \\ \Xhline{1.0pt} %
\end{tabular}}
\end{table}

\textbf{Impact of channel coding:} In LiDAR point cloud-based 3D object detection, the heterogeneous distribution characteristics of geometric and attribute information cause the detection heads to show different levels of sensitivity to distortions induced by channel noise. As shown in Table \ref{Table 7}, the proposed channel coder significantly enhanced transmission reliability across diverse channel conditions, which maintained detection accuracy.

\begin{table}[]
\setlength{\abovecaptionskip}{-0.1cm}
\caption{ABLATION STUDY OF POINT CLOUD UPSAMPLING TASK. THE BEST RESULTS ARE SHOWN IN BOLD.}
\resizebox{\columnwidth}{!}{
\label{Table 8}
\begin{tabular}{c|c|ccc|ccc}
\Xhline{1.0pt} %
\multirow{2}{*}{\begin{tabular}[c]{@{}c@{}}Dilated Sparse Convolution\\ Initialized Prompt\end{tabular}} & \multirow{2}{*}{\begin{tabular}[c]{@{}c@{}}Diffusion Model\\ based Point Cloud Generation\end{tabular}} & \multicolumn{3}{c|}{Car BEV AP (R40)}                                                            & \multicolumn{3}{c}{Car 3D AP (R40)}                                                              \\ \cline{3-8} 
                        &                                                                                                         & \multicolumn{1}{c|}{Easy}             & \multicolumn{1}{c|}{Moderate}         & Hard             & \multicolumn{1}{c|}{Easy}             & \multicolumn{1}{c|}{Moderate}         & Hard             \\ \Xhline{1.0pt} %
Original                                                                                            & Original                                                                                                & \multicolumn{1}{c|}{96.08}          & \multicolumn{1}{c|}{92.04}          & \textbf{91.44} & \multicolumn{1}{c|}{92.89}          & \multicolumn{1}{c|}{88.07}          & \textbf{85.49} \\
\ding{51}                                                                                                   & \ding{55}                                                                                                       & \multicolumn{1}{c|}{96.75}          & \multicolumn{1}{c|}{91.77}          & 86.70          & \multicolumn{1}{c|}{93.45}          & \multicolumn{1}{c|}{88.01}          & 80.63          \\
\ding{55}                                                                                                   & \ding{51}                                                                                                       & \multicolumn{1}{c|}{96.83}          & \multicolumn{1}{c|}{93.54}          & 86.01          & \multicolumn{1}{c|}{95.71}          & \multicolumn{1}{c|}{89.43}          & 81.99          \\
\ding{51}                                                                                                   & \ding{51}                                                                                                       & \multicolumn{1}{c|}{\textbf{96.95}} & \multicolumn{1}{c|}{\textbf{96.29}} & 88.76          & \multicolumn{1}{c|}{\textbf{96.19}} & \multicolumn{1}{c|}{\textbf{92.62}} & 85.13          \\ \Xhline{1.0pt} %

\end{tabular}
}
\end{table}

\textbf{Impact of prompt generation block:} For the feature point cloud upsampling task, we combined dilated sparse convolution with a diffusion model. To assess the individual contributions of these two components, we conducted an ablation study, and the results are reported in Table \ref{Table 8}. The findings show that combining the two components achieved the best overall performance, while removing either component caused a noticeable degradation in accuracy. Moreover, the introduction of dilated sparse convolution layers accelerated the denoising process in the diffusion model, enabling high-quality point cloud upsampling in just a single denoising step.
\vspace{-0.3cm} 
\subsection{System Efficiency Analysis}

\begin{table}[]
\setlength{\abovecaptionskip}{-0.1cm}
\caption{COMPUTATIONAL COMPLEXITY}
\resizebox{\columnwidth}{!}{
\label{Table 9}
\begin{tabular}{c|c|c|cc|cc}
\Xhline{1.0pt} %
\multirow{2}{*}{Scheme} & \multirow{2}{*}{Method} & \multirow{2}{*}{\begin{tabular}[c]{@{}c@{}}Channel\\ SNR (dB)\end{tabular}} & \multicolumn{2}{c|}{Latency (ms)}          & \multicolumn{2}{c}{Parameter (M)}         \\ \cline{4-7} 
                        &                         &                                                                             & \multicolumn{1}{c|}{Edge}      & Cloud     & \multicolumn{1}{c|}{Edge}    & Cloud      \\ \Xhline{1.0pt} %
1                       & VirConv-L               & ——                                                                          & \multicolumn{1}{c|}{124.37}  & 0         & \multicolumn{1}{c|}{13.20} & 0 \\
2                       & Proposed                & 20                                                                          & \multicolumn{1}{c|}{1483.50} & 192.55  & \multicolumn{1}{c|}{1.89}  & 13.61    \\
3                       & Proposed                & 15                                                                          & \multicolumn{1}{c|}{1483.50} & 194.68  & \multicolumn{1}{c|}{1.89}  & 13.61    \\
4                       & Proposed                & 10                                                                          & \multicolumn{1}{c|}{1483.50} & 198.17  & \multicolumn{1}{c|}{1.89}  & 13.61    \\
5                       & Proposed                & 5                                                                           & \multicolumn{1}{c|}{1483.50} & 234.62  & \multicolumn{1}{c|}{1.89}  & 13.61    \\
6                       & Proposed                & 0                                                                           & \multicolumn{1}{c|}{1483.50} & 5361.96 & \multicolumn{1}{c|}{1.89}  & 13.61    \\ \Xhline{1.0pt} %
\end{tabular}}
\end{table}

We evaluated the computational complexity of the proposed method in terms of latency and parameter count. As shown in Table \ref{Table 9}, the proposed method introduced only a marginal increase in parameters compared with the baseline model. Furthermore, as the channel condition deteriorated, the decoding latency increased progressively, since the LDPC decoder required more iterations to achieve convergence under poor channel conditions. In practical edge-cloud collaborative scenarios, the cloud device has significantly higher computational resources, thus the latency at the cloud device can be further reduced. 

\section{Conclusion}
We proposed an object detection-driven point cloud compression network that simultaneously meets the requirements of efficient compression and reliable transmission. Built upon the multimodal detector VirConv-L, the proposed approach overcomes the limitations of using images alone, which lack depth information, and point clouds alone, where distant points are relatively sparse. Our work introduces three key innovations: a foreground prediction and attention-based channel compaction strategy that selectively compresses task-relevant regions, reducing transmission volume while enhancing detection accuracy; an unequal-protection-based channel coding scheme that dynamically adapts feature compression to channel conditions, significantly improving robustness against channel errors compared with direct transmission methods; and a diffusion model-based feature point cloud upsampling scheme that generates high-resolution task-relevant features from reconstructed low-resolution representations, substantially reducing data transmission overhead. Extensive experiments on the KITTI dataset demonstrate that the proposed framework achieves state-of-the-art compression efficiency, improves detection accuracy for easy and moderate objects with negligible loss for hard objects, and maintains strong generalization across diverse channel conditions. Future work will investigate unified compression strategies to support a wider range of perception tasks, including semantic segmentation and object tracking.

\footnotesize
\bibliographystyle{IEEEbib}
\bibliography{IEEE-Transactions-LaTeX2e-templates-and-instructions/ref}
\vspace{-0.5cm}

\begin{IEEEbiographynophoto}{Chongzhen Tian}
received the B.S. degree and M.S. degree from Ningbo University, Ningbo, China, in 2020 and 2023, respectively. He is currently pursuing the Ph.D. degree at the Shandong University, Jinan, China. His research interests include point cloud compression and quality assessment.
\end{IEEEbiographynophoto}

\begin{IEEEbiographynophoto}{Hui Yuan} (\textit{Senior Member, IEEE}) received the B.E. and Ph.D. degrees in telecommunication engineering from Xidian University, Xi’an, China, in 2006 and 2011, respectively. In April 2011, he joined Shandong University, Ji’nan, China, as a Lecturer (April 2011–December 2014), an Associate Professor (January 2015-August 2016), and a Professor(September 2016). His current research interests include 3D visual media coding, processing, and communication.
\end{IEEEbiographynophoto}

\begin{IEEEbiographynophoto}{Pan Zhao} received the B.E. degree from the School of Software Engineering, Jinling Institute of Technology, Nanjing, China, in 2021, and the M.S. degree from School of Artificial Intelligence, Nanjing University of Information Science and Technology,
Nanjing, China, in 2024. He is currently pursuing the Ph.D. degree with the School of Control Science and Engineering, Shandong University, Jinan, China. His research interests include point cloud compression and quality enhancement.
\end{IEEEbiographynophoto}

\begin{IEEEbiographynophoto}{Chang Sun} received the B.S. and M.S. degrees in school of control science and engineering from Shandong University, Shandong, China, in 2019 and 2022. He is currently working toward the Ph.D degree in Shandong University. His research interests include point cloud compression and processing.
\end{IEEEbiographynophoto}

\begin{IEEEbiographynophoto}{Raouf Hamzaoui} (\textit{Senior Member, IEEE}) received the M.Sc. degree in mathematics from the University of Montreal, Montreal, QC, Canada, in 1993, the Dr.rer.nat. degree from the University of Freiburg, Freiburg im Breisgau, Germany, in 1997, and the Habilitation degree in computer science from the
University of Konstanz, Konstanz, Germany, in 2004. He was an Assistant Professor with the Department of Computer Science, University of Leipzig, Leipzig, Germany, and Department of Computer and Information Science, University of Konstanz. In 2006, he joined De Montfort University, Leicester, U.K., where he is currently a Professor in media technology
\end{IEEEbiographynophoto}

\begin{IEEEbiographynophoto}{Sam Kwong} (\textit{Fellow, IEEE}) is currently the Chair Professor of computational intelligence, and concurrently as Associate Vice-President (Strategic Research) of Lingnan University, Hong Kong. He is a Distinguished Scholar of evolutionary computation, artificial intelligence (AI) solutions, and image/video processing, with a strong record of scientific innovations and real-world impacts. 		
\end{IEEEbiographynophoto}


\end{document}